\documentclass[a4paper,fleqn,usenatbib]{mnras}

\usepackage{graphicx}	
\usepackage[autostyle]{csquotes}
\usepackage{changepage}
\usepackage{amsmath}	
\usepackage{amssymb}	
\usepackage{multicol}
\usepackage{multirow}
\usepackage{booktabs}
\usepackage[flushleft]{threeparttable}

\title[SEDs of 6.7 GHz methanol masers]{Probing the early phases of high mass star formation with 6.7 GHz methanol masers}

\author[Paulson \& Pandian]{
Sonu Tabitha Paulson,$^{1}$\thanks{E-mail: sonutabitha.15@res.iist.ac.in}
Jagadheep D. Pandian$^{1}$\thanks{E-mail: jagadheep@iist.ac.in }
\\
$^{1}$Department of Earth and Space Sciences, Indian Institute of Space Science \& Technology, Thiruvananthapuram 695547, Kerala, India\\
}

\date{Accepted 2019 December 12. Received 2019 December 10; in original form 2019 May 8}

\pubyear{2019}

\begin{document}
\label{firstpage}
\pagerange{\pageref{firstpage}--\pageref{lastpage}}
\maketitle

\begin{abstract}
Methanol masers at 6.7 GHz are the brightest of class II methanol masers and have been found exclusively towards massive star forming regions. These masers can thus be used as a unique tool to probe the early phases of massive star formation. We present here a study of the spectral energy distributions of 320 6.7 GHz methanol masers chosen from the MMB catalogue, which fall in the Hi-GAL range ($|l| \le 60^{\circ}$, $|b| \le 1^{\circ}$). The spectral energy distributions are constructed from $870 - 70~\mu$m using data from the ATLASGAL and Hi-GAL surveys. The emission from cold dust is modelled by a single grey body component fit. We estimate the clump properties such as mass, FIR luminosity and column density using the best fit parameters of the SED fits. Considering the Kauffman criteria for massive star formation, we find that all but a few maser hosts have the potential to harbour at least one high mass star. The physical properties of the methanol maser hosts are also discussed. The evolutionary stages of 6.7 GHz maser sources, explored using the mass luminosity diagram, suggests that they are predominantly associated with high mass stars with the majority being in the accretion phase. However, we observe a small number of sources that could possibly be related to intermediate or low-mass stars.
\end{abstract}


\section{INTRODUCTION}

Massive stars ($M>8M_{\odot}$) are dominant players in the galaxy. These stars have a profound impact on their local environment through massive outflows, strong stellar winds, UV radiation and supernovae explosions. They chemically enrich the interstellar medium as they are the principal sources of heavy elements. In spite of the important role that the massive stars play in shaping the evolution of the Universe, the full paradigm of their formation and early evolution is poorly understood. This is mainly due to the fact that the high-mass star formation is a rapid process that takes place while the stars are deeply embedded in dense clumps. Other observational challenges include their rarity, relatively large distance scales and the clustered environment in which they form.

A common phenomenon observed in active high-mass star forming regions is the presence of interstellar masers. While water masers are found in both low-mass and high-mass star forming regions, and OH masers are found towards both high-mass star forming regions and evolved stars, Class II methanol masers (the brightest transition being at 6.7~GHz) are found exclusively towards high-mass star forming regions (e.g. \citealt{xu2008high}, \citealt{breen2013confirmation} and references therein). This makes 6.7 GHz methanol masers a unique tool to detect and probe the early phases of massive star formation. Since the first detection reported by \citet{menten1992vlbi}, there have been numerous targeted and blind surveys for 6.7~GHz methanol masers, the largest being the Methanol Multibeam survey (MMB; \citealt{caswell20106}).

One of the key questions concerning 6.7~GHz methanol masers is whether or not they are indeed exclusively associated with massive star formation. Targeted searches for 6.7~GHz methanol masers towards low-mass young stellar objects and hot corinos have not yielded any detections \citep{minier2003protostellar, bourke2005identification,xu2008high,pandian2008detection}. \citet{minier2003protostellar} give a lower limit of 3~M$_\odot$ to the mass of a star that is associated with bright (brightness temperature $> 3 \times 10^6$~K) 6.7~GHz methanol masers. \citet{bourke2005identification} find the lower luminosity limit for a source to host 6.7~GHz methanol masers to be around $10^3$~L$_\odot$. Several targeted studies of 6.7~GHz methanol masers based on spectral energy distributions and infrared colors (\citealt{pandian2010spectral}, \citealt{ellingsen2006methanol} and references therein) also conclude the association of 6.7~GHz methanol masers with the early phases of massive star formation. \citet{urquhart2013atlasgal} studied the properties of 577 ATLASGAL \citep{schuller2009atlasgal} clumps associated with 6.7~GHz methanol masers and concluded that over 90\% of their sample are associated with massive young stars. However, they also concluded that a small number of 6.7~GHz methanol masers may be associated with clumps that may form only intermediate-mass stars. While the study of \citet{urquhart2013atlasgal} covers a signficant fraction of the 6.7~GHz methanol masers that have been detected using the Methanol Multibeam Survey, it used only the 870~$\mu$m ATLASGAL data and derived the properties of the clumps hosting methanol masers assuming the dust temperature to be 20~K.

Another key question concerns the evolutionary stage of the young stellar objects that excite 6.7~GHz methanol masers. Based on the spectral energy distributions of 6.7~GHz methanol masers, \citet{pandian2010spectral} conclude that the masers are associated with rapidly accreting massive stars, mostly prior to the formation of an ultracompact H~\large ii (UCH~\large ii) region. However, this study is limited by the small sample size, and the lack of data in the far-infrared. \citet{de20156} studied the association of 6.7~GHz methanol masers with molecular outflows traced by $^{13}$CO and concluded that the masers turn on in hot core sources that have already developed outflows and turn off during the UCH~\large ii phase.

While earlier studies such as \citet{urquhart2013atlasgal} and \citet{pandian2010spectral} were constrained by the lack of data in the far-infrared, the availability of data from the  Herschel Infrared Galactic Plane Survey (Hi-GAL; \citealt{molinari2010hi}) allows us to do a more systematic study of the SEDs of the sources exciting methanol masers. In this paper, we present the SEDs of 320 6.7~GHz methanol masers from 870~$\mu$m to 70~$\mu$m using data from ATLASGAL and Hi-GAL surveys. 

The structure of the paper is as follows. In section~\ref{sec2}, we discuss the source selection and data analysis. The results of our study are presented in section~\ref{sec3}. A discussion of the results in the larger context of the relation between 6.7~GHz methanol masers and massive star formation is presented in section~\ref{sec4}, and we summarise our results in section~\ref{sec5}.

\begin{figure*}
\centering
\includegraphics[width=\textwidth]{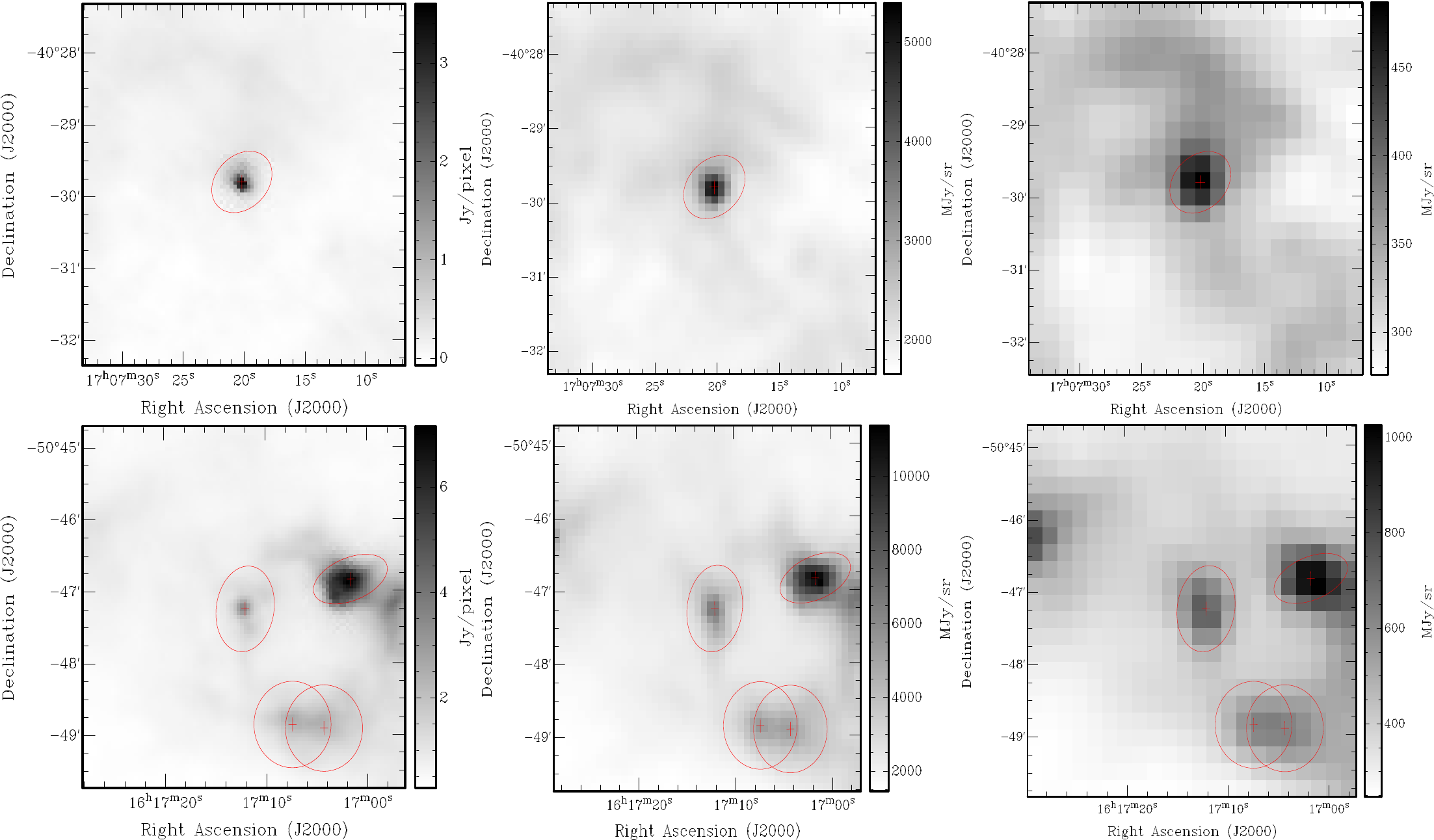}
\caption{Source identification and apertures used by \textsc{\large{hyper}} for doing photometry. The two rows show the 160, 250 and 500 ~$\mu$m data for G346.036+0.048 and G332.560$-$0.148 respectively, the latter showcasing a slightly crowded field.}
\label{fig:hypereg}
\end{figure*}

\begin{table*}
\caption{Flux densities of sources hosting 6.7~GHz methanol masers from 870 to 70~$\mu$m. The uncertainties in the measured flux densities are given in parenthesis.}
         \label{fluxtable}
\begin{center}
         \centering
         \begin{threeparttable}
           \begin{tabular}{ccccccccc}
    \hline
    \hline
    \multirow{2}{*}{Name} &\multicolumn{7}{c}{Flux Values (Jy)} \\
      && {870~$\mu$m} & {500~$\mu$m} & {350~$\mu$m} & {250~$\mu$m} & {160~$\mu$m} & {70~$\mu$m} \\
      \midrule
    G345.131-0.174  &&4.14 (0.11)  &17.52 (0.77)  &49.45 (1.56)  &97.06 (2.64)  &125.80 (4.46)  &60.30 (2.92) \\
    G345.576-0.225  &&1.90 (0.06)  &6.71 (0.40)  &19.95 (0.80)  &29.80 (1.02)  &33.28 (0.90)  &3.85 (0.30)  \\
    G345.807-0.044  && 0.84 (0.05) &3.92 (0.32)  &8.03 (0.48)  &11.70 (0.65)  &9.75 (0.48)  &0.54 (0.15) \\
    G345.824+0.044  &&2.85 (0.09)  &5.40 (0.82)  &21.25 (2.15)  &44.66 (2.92)  &61.32 (3.30)  &44.96 (2.74)  \\
    G309.384-0.135  &&7.22(0.29)  &34.35 (1.04)  &104.22 (2.65)  &210.16 (5.33)  &376.29 (11.26)  &270.82 (7.67)  \\
    G5.618-0.082  &&4.31 (0.15)  &20.19 (0.48)  &53.77 (1.19)  &94.64 (2.18)  &98.20 (3.26)  &17.61 (0.33) \\
    G16.855+0.641  &&1.80 (0.06)  &8.25 (0.37)  &23.47 (0.78)  &40.89 (1.05)  &55.24 (1.68)  &18.91 (0.69) \\
    G010.724$-$0.334  &&4.05 (0.16)  &18.46 (0.98)  &55.74 (1.99)  &81.11 (2.97)  &112.08 (3.70) &39.01 (1.67) \\
    
    \hline
  \end{tabular}
  \begin{tablenotes}
            \item[]*The full table will be made available online.
            
        \end{tablenotes}
  \end{threeparttable}
   \end{center}
\end{table*}
\section{SOURCE SELECTION AND DATA ANALYSIS}\label{sec2}
\begin{figure*}
\begin{center}

\includegraphics[width=0.44\textwidth, trim= 0 0.4cm 0 0]{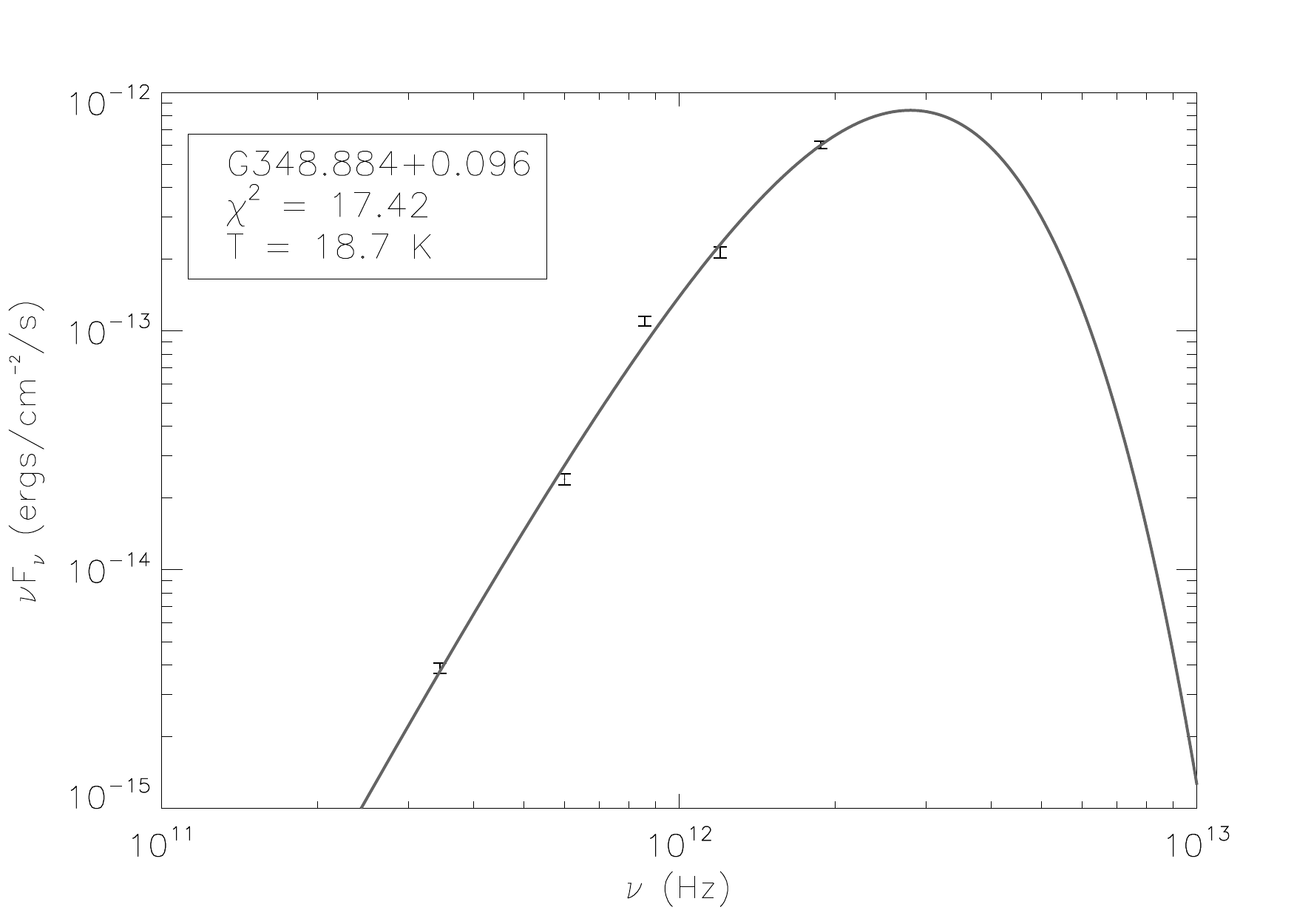}
\includegraphics[width=0.44\textwidth, trim= 0 0.4cm 0 0]{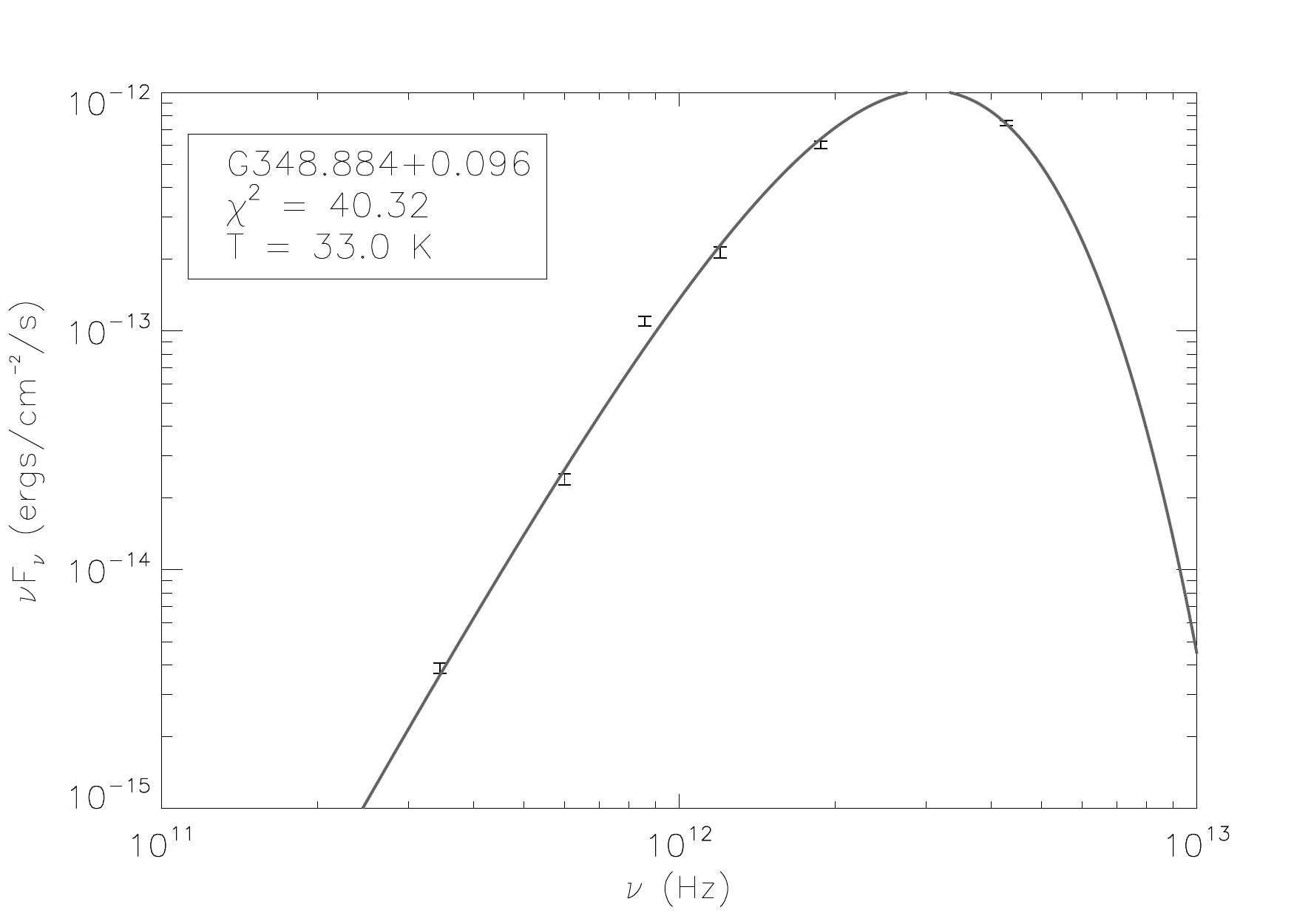}
\includegraphics[width=0.44\textwidth, trim= 0 0.4cm 0 0]{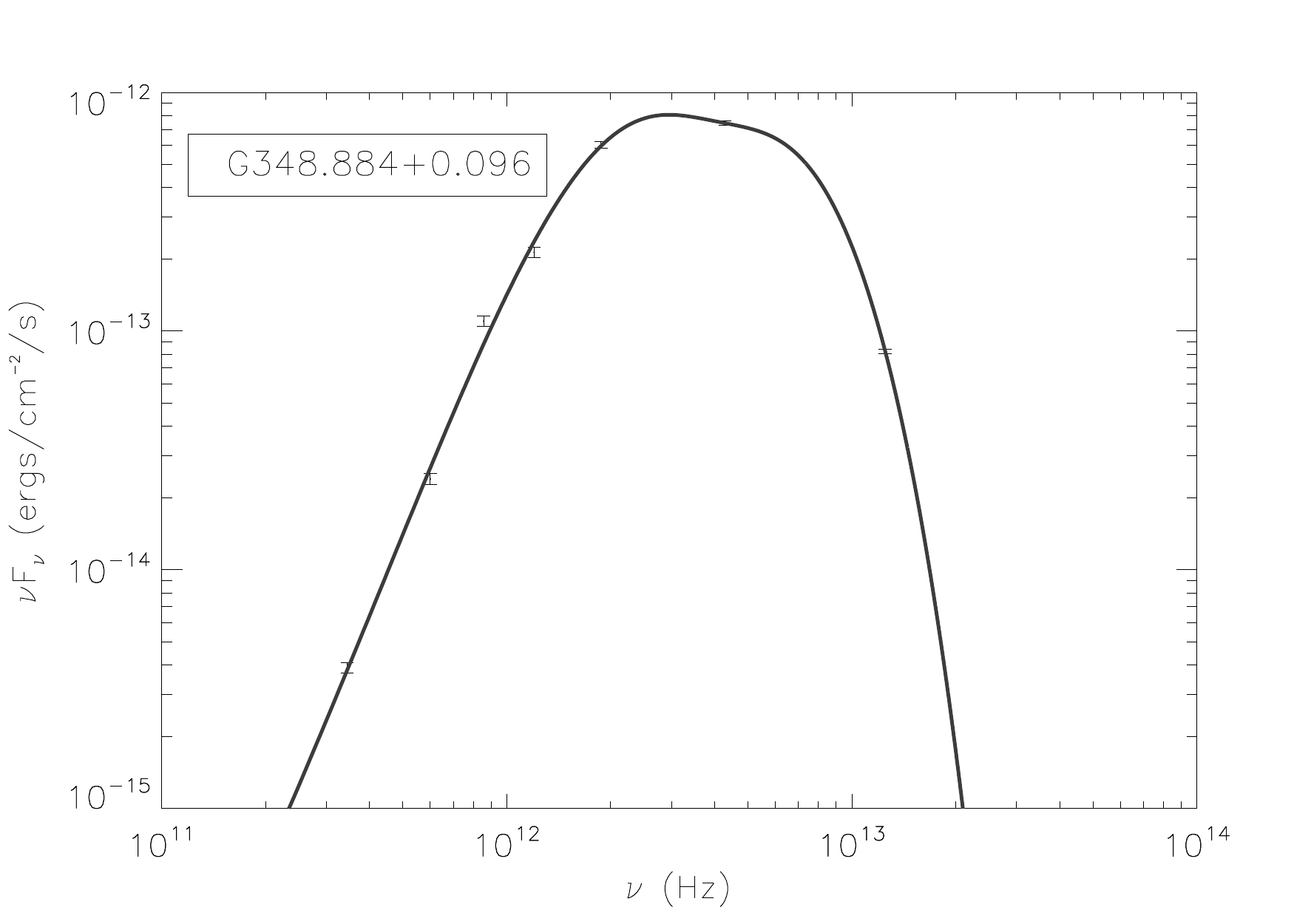}
\caption{\label{fig:test} SED fits at different wavelength ranges.The upper left panel: The grey body fit for the wavelength range 70-870~$\mu$m. The upper right panel: grey body fit for the wavelength range 160-870~$\mu$m. Lower panel: SED fit taking into account both the cold and warm dust emission (excluding the contribution of 70~$\mu$m towards the cold dust emission).}

\end{center}

\end{figure*}

Our methanol maser sample has been selected from the catalog of the Methanol Multibeam survey. The MMB survey covers the entire Galactic plane with a latitude coverage of $|b|$ \ensuremath{\le} $2^{\circ}$ and the catalog comprises of a total of 972 sources. We have restricted our analysis to sources that have been covered by the Hi-GAL survey ($|l|\leq60^{\circ}$, $|b|\le1^{\circ}$). This limits the number of masers in our sample to 630. Determination of physical parameters such as mass, size and luminosity of the sources requires a knowledge of the distances to the sources; and this brings the sample size to 602 sources. The distances are taken from \citet{urquhart2013atlasgala}, \citet{reid2014trigonometric}, \citet{pandian2009arecibo} and \citet{green20176}. The distances for a majority of the sources are determined from their kinematics using the observed radial velocity, with a smaller sample of sources having more accurate distance estimates through trigonometric parallax. \citet{green20176} also report distances to the full MMB sample using a parallax based approach of \citet{reid2016parallax} wherein sources are assigned to spiral arms using a Bayesian approach. However, for this work we prefer to use the kinematic distances since a large fraction of our sample is in the fourth Galactic quadrant where parallax measurements are limited. The method of \citet{reid2016parallax} is expected to be more accurate in the northern hemisphere where deviations in the kinematics of the star forming regions from pure Galactic rotation are modelled using observed parallax distances. An examination of the new distances reported by \citet{green20176} indeed shows a large number of sources to be not associated with any spiral arm. A discussion on the variation of the results when adopting the new parallax based distances is presented in section 4.4.

We have discarded sources that are either saturated in the Hi-GAL data or are in very crowded fields where reliable photometry is difficult (especially at 500~$\mu$m where source blending in such fields is severe). We have also taken into account only sources that have ATLASGAL counterparts. This gave a final methanol maser sample of 320 sources among which the integrated flux densities of 311 methanol maser sources are taken from \citet{breen20156}. 
 
We have determined the spectral energy distributions from 870~$\mu$m to 160~$\mu$m of the sources hosting methanol masers using data from ATLASGAL and Hi-GAL surveys. Fluxes from 870 to 70~$\mu$m have been determined from aperture photometry of the ATLASGAL and Hi-GAL images (level 2.5 images in the Herschel archive).  The 870~$\mu$m ATLASGAL data have a resolution of 19$\arcsec$, while the Hi-GAL data have a resolution of 36.9$\arcsec$, 25.3$\arcsec$, 18.0$\arcsec$, 11.6$\arcsec$ and 10.2$\arcsec$ at 500, 350, 250, 160 and 70~$\mu$m respectively.

\subsection{Source Photometry}
Aperture photometry of Hi-GAL data is extremely challenging due to the complex structure of background emission and the crowded fields. An additional complication is that the resolution is different at different wavelength bands. We carried out aperture photometry using \textsc{\large{hyper}}, an IDL based software package \citep{traficante2015hyper}. \textsc{\large{hyper}} first detects sources in each band based on the supplied threshold. The photometry is then carried out using the sources detected in the reference band that is specified by the user. First, small cutouts are made around each source. After masking the central pixels containing emission from the source, the background is estimated by fitting a two-dimensional polynomial up to fourth order. The order of the polynomial is decided such that the residual after subtraction of the background is minimized. In the case of crowded fields where emission from multiple sources overlap, the size of the cutout is chosen to include all overlapping sources. A more detailed description of the procedure can be found in \citet{traficante2015hyper}.

The shape of the aperture used for performing photometry is elliptical and is determined by fitting a two-dimensional Gaussian to the data at the reference wavelength. In the case of blended sources, a simultaneous multi-Gaussian fit is done to separate between the target source and its companions. The size of the aperture is related to the measured size of the source by a user supplied factor. This factor is selected such that most of the flux is recovered while minimizing contamination from nearby sources in crowded fields. The same aperture is then used to determine the fluxes in all bands. This allows us to obtain the integrated flux from the same volume of gas and dust at different wavelengths. We have used 250~$\mu$m as a reference wavelength for aperture photometry since the resolution and morphology of emission at 250~$\mu$m is very similar to that at 870~$\mu$m while the signal to noise ratio is much better. Fig.~\ref{fig:hypereg} shows an example of photometry for an isolated and a slightly crowded field.

It is to be noted that the aperture photometry above will not recover the total flux from the source, especially at 350 and 500~$\mu$m, unless the size of the aperture is very large. However, the use of a large aperture factor is detrimental to the quality of photometry due to the effect of residuals from the background subtraction and contamination from other sources in crowded fields. Hence, the scale factor between the size of the source and that of the aperture used for photometry is kept to a moderate value, and the total flux is determined from the flux obtained by \textsc{\large{hyper}} by incorporating a wavelength dependent correction factor. The correction factor is determined by simulating sources of different sizes, convolving the simulated data to the resolution of the Hi-GAL and ATLASGAL surveys, and comparing the actual flux with those determined by \textsc{\large{hyper}}.

Table~\ref{fluxtable} shows the flux densities of the sources from 870~$\mu$m to 70~$\mu$m. The uncertainties in the flux densities quoted in Table~\ref{fluxtable} only account for random errors in the data and do not incorporate systematic effects such as the accuracy of flux calibration in the ATLASGAL and Hi-GAL surveys. A small fraction ($< 10\%$) of the sources are resolved into multiple sources at 160 and 70~$\mu$m even though they appear as single sources at 250~$\mu$m. In these cases, the procedure of aperture photometry used by \textsc{\large{hyper}} will determine the combined fluxes of the multiple sources at shorter wavelengths since the aperture ellipse used for photometry is identical to that at 250~$\mu$m. The same result would have been obtained if the data at shorter wavelengths were convolved to the same resolution as that of the reference wavelength.

Another point to note is that there are alternate algorithms for carrying out aperture photometry in the Galactic plane region. While a detailed list of these algorithms can be found in \citet{traficante2015hyper}, it is of interest to compare our results with that of the \textsc{\large{cutex}} algorithm, which is used for photometry of the catalogue released by the Hi-GAL team (e.g. \citealt{elia2017hi}). While the \textsc{\large{cutex}} algorithm is relatively insensitive to background contamination for identification of point sources, the photometry is done independently in each band by fitting the source with a 2D Gaussian. Since the beam size of Herschel changes by more than a factor of 4 over the different bands, the process of aperture photometry requires integrating flux from larger areas at long wavelengths. This results in a bias which is corrected by rescaling the fluxes according to the ratio of deconvolved source sizes using 250~$\mu$m as a reference \citep{elia2017hi}. A comparison of our fluxes with those of \citet{elia2017hi} shows consistency to within 15\% although the 500~$\mu$m fluxes show a mean variation of 30\%. In most cases where there is significant variation between the \textsc{\large{hyper}} and \textsc{\large{cutex}} fluxes, the source sizes are seen to be significantly different. Moreover, the source sizes are seen to be significantly larger at 500~$\mu$m compared to shorter wavelengths. Hence, we conclude that the discrepancy is mostly due to blending of sources due to the relatively poor resolution at 500~$\mu$m. This highlights the limitation of the \textsc{\large{cutex}} algorithm due to its treatment of the different bands independently.

\subsection{Fitting the SED}

We have fit the SED of the sources from 870~$\mu$m to 160~$\mu$m using a grey body to model the cold dust emission:
\begin{align}
F_{\nu} & = \Omega_{c}B_{\nu}(T_c)(1-e^{-\tau_{\nu}}) \\
\tau_{\nu} & = \tau_{0}\left(\frac{\nu}{\nu_{0}}\right)^{\beta} 
\end{align}
Here, $\Omega_{c}$ is the deconvolved solid angle of the source derived from the aperture used by \textsc{\large{hyper}} for doing photometry (deconvolution is done using the Hi-GAL beam size at 250~$\mu$m -- i.e. 18$''$), $T_c$ is the temperature of the cold dust, $\tau_0$ is the optical depth at frequency $\nu_0$ (chosen to be 500~$\mu$m wavelength) and $\beta$ is the dust emissivity index which is assumed to be $1 < \beta < 3$. The fitting was carried out using the nonlinear least squares Marquardt-Levenberg algorithm.

The emission at 70~$\mu$m is not likely to be attributable to a single dust component \citep{compiegne2010dust,battersby2011characterizing}. Rather, the 70~$\mu$m flux is expected to have contributions from both cold and warm dust. In order to test this hypothesis, we first compare the single component fit to the SED of a source with and without including the 70~$\mu$m flux. As shown in the left and right panels of Fig.~\ref{fig:test} for a typical source (G348.884+0.096 in this case), the fit including the 70~$\mu$m data gives a higher dust temperature with a significantly poorer $\chi^2$ goodness of fit. We then carried out a two component fit to the SED including the the 24~$\mu$m flux from the MIPSGAL catalogue \citep{gutermuth2014mipsgal}, with the second component being modelled as a black body:
\begin{equation}
F_{\nu}=\Omega_{w}B_{\nu}(T_w)
\end{equation}
where $\Omega_w$ and $T_w$ are the solid angle and the temperature of the warm dust respectively. In this fit (bottom panel of Fig.~\ref{fig:test}), almost 60$\%$ of the 70~$\mu$m flux arose from the warm dust emission, confirming the hypothesis of the 70~$\mu$m emission arising from both cold and warm dust. Hence, for most sources, the grey body fit for cold dust was restricted to the 870$-$160~$\mu$m data.However, there are a few sources where the temperature of the cold dust is high enough that the peak of the black body is at wavelengths significantly shorter than 160~$\mu$m. In such cases, it is not possible to determine the dust temperature by fitting the 870$-$160~$\mu$m data alone since these wavelengths lie in the Rayleigh-Jeans part of the spectrum. In such cases, we have included the 70~$\mu$m data to fit the SED.

\section{RESULTS}\label{sec3}
\begin{figure*}
\centering
 \includegraphics[scale=0.8]{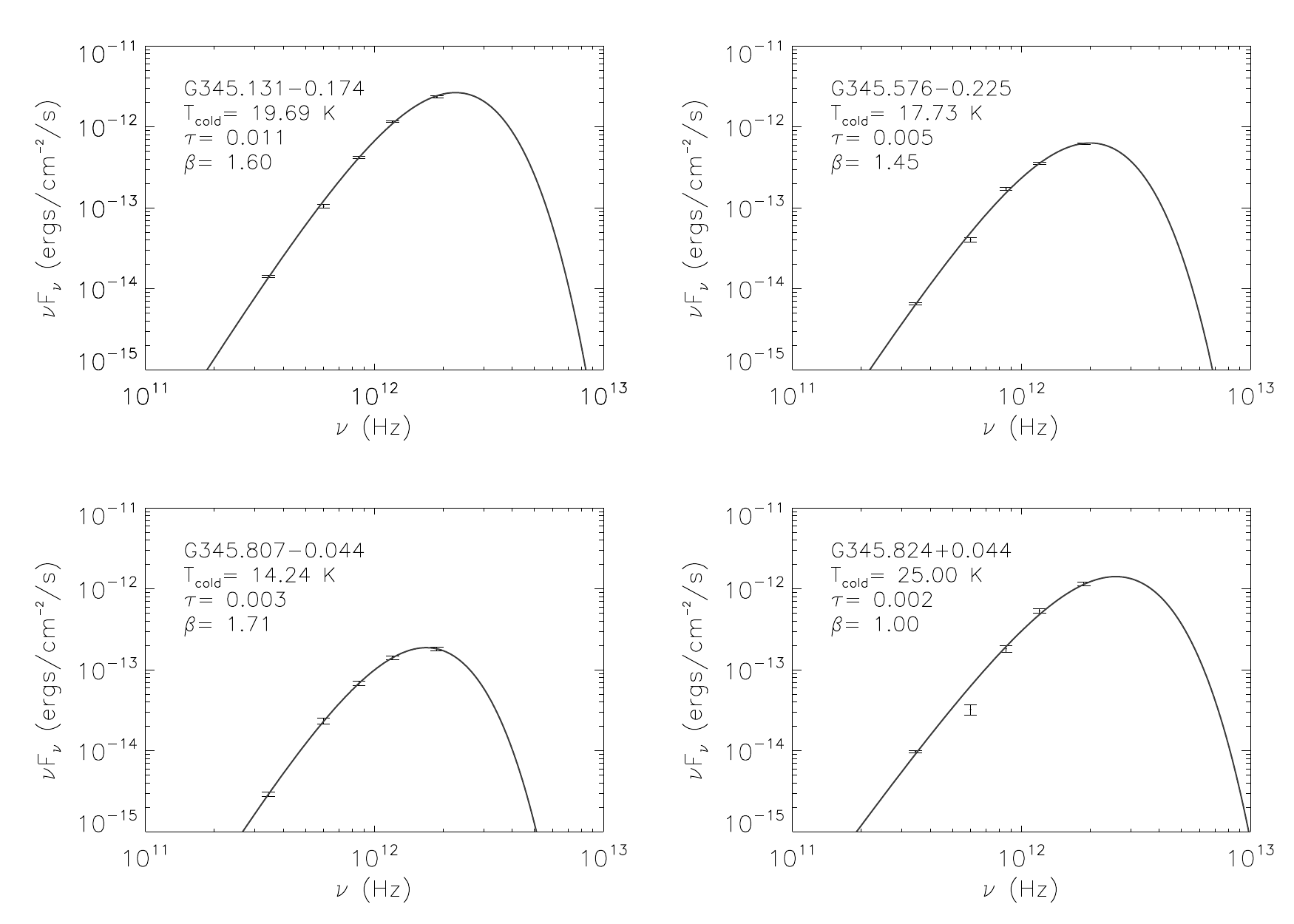}
\caption{Single component fits to the SEDs of characteristic sources. The inset shows the source name, temperature of cold dust, dust optical depth at 500~$\mu$m and the dust spectral index.}
\label{fig:samplefits}
\end{figure*}

We have fit the cold dust emission for 318 sources from 870$-$160~$\mu$m and 2 sources from 870$-$70~$\mu$m. The fits to SEDs of characteristic sources are shown in Fig.~\ref{fig:samplefits} and the fit parameters for the sources listed in Table~\ref{fluxtable} are shown in Table~\ref{Bestfitparams}. The uncertainty in the temperature of cold dust from the SED fit is typically less than 15\%. The properties of the clumps hosting methanol masers such as radii, hydrogen column densities, masses, surface densities and luminosities were determined from the SED parameters. These properties are described in subsections below. Table~\ref{Bestfitprops} lists the physical properties of each clump, while the global statistics are summarized in Table~\ref{param_stats}.

\begin{table*}
\caption{The best fit parameters of characteristic sources. The columns show the source name, distance, temperature of the cold dust component, the solid angle of cold dust as seen in aperture photometry, optical depth of dust at 500~$\mu$m and the dust spectral index, $\beta$.}
         \label{Bestfitparams}
\begin{center}
         \centering
         \begin{threeparttable}
  \begin{tabular}{cccccc}
    \hline
    \hline
     Name & {Distance (kpc)} & {$T_{c}$ (K)} & {$\Omega_c$ (sr)} & {$\tau$} &{$\beta$}   \\
      \midrule
    G345.131$-$0.174  & 3.05 & 19.69 & $8.54 \times 10^{-9}$ & 0.024 & 1.6 \\
    G345.576$-$0.225  & 5.5 & 17.79  & $1.05 \times 10^{-8}$ & 0.010 & 1.5 \\
    G345.807$-$0.044  & 10.8 & 14.29 & $1.13 \times 10^{-8}$ & 0.006 & 1.7 \\
    G345.824$+$0.044  & 10.9 & 25.04 & $1.36 \times 10^{-8}$ & 0.005 & 1.0 \\
    G005.618$-$0.082  & 5.1  & 15.06 & $6.15 \times 10^{-9}$ & 0.092 & 2.0 \\
    G010.724$-$0.334  & 5.2  & 18.47 & $7.25 \times 10^{-9}$ & 0.003 & 1.7 \\
    G016.855$+$0.641  & 13.79  & 17.40 & $3.04 \times10^{-9}$ & 0.049 & 1.8 \\
    G309.384$-$0.135  & 5.4 & 24.24 & $6.90 \times 10^{-9}$  & 0.039 & 1.6 \\

    \hline
  \end{tabular}
  \begin{tablenotes}
            \item[] *The full table will be made available online.
            
        \end{tablenotes}
  \end{threeparttable}
  \end{center}
\end{table*}

\begin{table*}
\caption{The physical properties derived from best fit parameters of characteristic sources. The columns show the source name, clump Mass, Hydrogen column density,effective radius,surface density and  FIR Luminosity.}
         \label{Bestfitprops}

\begin{center}
         \centering
         \begin{threeparttable}
  \begin{tabular}{cccccc}
    \hline
    \hline
     Name & {Clump Mass (M$_{\odot}$)} & {$N_{H_2}$ (cm$^{-2}$)} & {Radius (pc)} & {$\Sigma$ (g cm$^{-2}$)} & {L$_\mathrm{FIR}$ (L$_\odot$)}    \\
      \midrule
G345.131$-$0.174 & $2.13 \times 10^{2}$ & $1.20 \times 10^{24}$	& 0.307	& 0.150	& $8.42 \times 10^{2}$\\
G345.576$-$0.225 & $3.77 \times 10^{2}$	& $5.30 \times 10^{23}$	& 0.618	& 0.065	& $6.68 \times 10^{2}$\\
G345.807$-$0.044 & $9.18 \times 10^{2}$	& $3.11 \times 10^{23}$	& 1.321	& 0.035	& $7.45 \times 10^{2}$\\
G345.824$+$0.044 & $1.35 \times 10^{3}$	& $3.75 \times 10^{23}$	& 1.508	& 0.039	& $6.192 \times 10^{3}$\\
G005.618$-$0.082 & $1.04 \times 10^{3}$	& $2.91 \times 10^{23}$	& 0.333	& 0.622	& $1.45 \times 10^{3}$\\
G010.724$-$0.334 & $6.79 \times 10^{2}$	& $1.55 \times 10^{24}$	& 0.462	& 0.210	& $1.95 \times 10^{3}$\\
G016.855$+$0.641 & $2.53 \times 10^{3}$	& $1.95 \times 10^{24}$	& 0.549	& 0.554	& $6.53 \times 10^{3}$\\
G309.384$-$0.135 & $8.82 \times 10^{2}$	& $1.96 \times 10^{24}$	& 0.450	& 0.288	& $9.51 \times 10^{3}$\\

    \hline
  \end{tabular}
  \begin{tablenotes}
            \item[] *The full table will be made available online.
            
        \end{tablenotes}
  \end{threeparttable}
  \end{center}
\end{table*}

\begin{table*}
\caption{Summary of derived parameters}
        \label{param_stats}

\begin{center}
  \centering
  \begin{tabular}{cccccc}
    \hline
    \hline
     Parameter & {Mean}  & {Standard deviation} & {Median} & {Min} & {Max}   \\
      \midrule
    Dust temperature (K) & 21.54 & 5.29 & 20.86 & 10.86 & 47.59 \\
    Dust emissivity & 1.83 & 0.29 & 1.83 & 1.00 & 2.85 \\
    Effective radius (pc)  & 0.69 & 0.41 & 0.61 & 0.07 & 1.99 \\
    Surface Density (g cm$^{-2}$) & 0.27 & 0.38 & 0.17 & 0.02 & 4.28 \\   
    Clump Mass (M$_\odot$)  & $1.57 \times 10^{3}$ & $2.46 \times 10^{3}$  & $9.30 \times 10^{2}$ & 11.07 & $1.62 \times 10^{4}$\\
    Column Density (cm$^{-2}$)  & $1.68 \times 10^{24}$ & $2.29 \times 10^{24}$
 & $1.15 \times 10^{24}$ & $7.95 \times 10^{22}$ & $2.50 \times 10^{25}$\\
    FIR luminosity (L$_{\odot}$) & $1.74 \times 10^4$ & $3.91 \times 10^4$ & $6.44 \times 10^3$ & $1.33 \times 10^2$ & $3.28 \times 10^5$ \\
    \hline
  \end{tabular}
  \end{center}
\end{table*}

\subsection{Source sizes}
\begin{figure}
\centering
\includegraphics[width=0.45\textwidth, trim = 0 0.4cm 0 0]{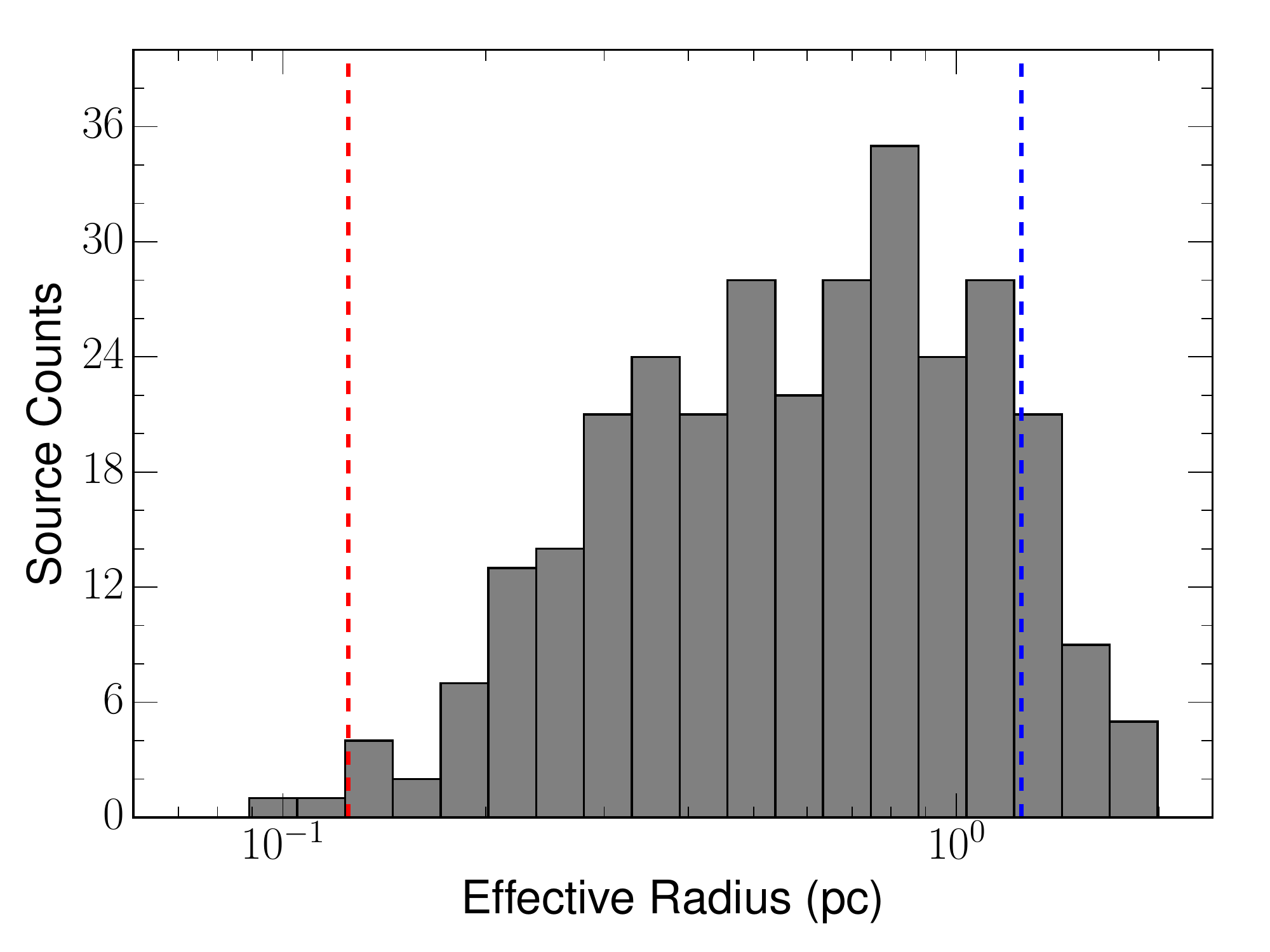}
\caption{Distribution of source size. The \textit{dashed red and blue} lines show nominal boundaries between cores and clumps, and clumps and clouds respectively.}\label{fig:sizedistribution}
\end{figure}

The effective radius of the sources have been estimated following the formulation of \citet{rosolowsky2010bolocam}:
\begin{equation}
\theta_{R} = \eta\big[{\big(\sigma_{maj}^{2}-\sigma_{beam}^{2}\big)\,\big(\sigma_{min}^{2}-\sigma_{beam}^{2}}\big)\big]^{1/4}
\end{equation}
where $\sigma_{beam}$ is related to the FWHM $\theta_{beam}$ by $\sigma_{beam} = \theta_{beam}/\sqrt{8\log2}$. Since the aperture photometry is carried out using 250~$\mu$m as the reference wavelength, $\theta_{beam}$ is the FWHM size of the Hi-GAL beam at 250~$\mu$m (i.e. 18.0$\arcsec$). The factor $\eta$ accounts for the relationship between rms size of the emission distribution and angular radius of the object, and is taken to be 2.4 as in \citet{rosolowsky2010bolocam}. Excluding sources that are close to being unresolved (for which deconvolution is not meaningful), the effective radius ranges from 0.07 to 1.99~pc, the median value being 0.69~pc. The distribution of the source radius is shown in Fig.~\ref{fig:sizedistribution}. \citet{bergin2007cold} suggest the nominal boundary between cores and clumps to be 0.125~pc and the boundary between clumps and clouds to be 1.25~pc. Under this nomenclature, three methanol masers are in cores, 273 are in clumps, and 33 are in clouds. However, all our data is from single dish telescopes, and it is likely that many sources may fragment into multiple objects when observed at higher angular resolution.

We find no correlation between the angular size of the source and its distance. This leads to a strong correlation between the source radius and distance to the source. Hence, the distinction between the identification of the source as a cloud, clump or core is primarily a result of the distance to the source. For instance, the sources that are classified as clouds are seen to have distances between 8.4 and 17.1~kpc with a mean value of 12.5~kpc, which is much larger than the mean distance to the entire sample (8.0 kpc). We will hence refer to all sources in our sample as clumps that will give rise to clusters rather than a single star. Our observation of lack of correlation between angular size of the source and the distance is similar to the findings of \citet{tackenberg2013triggered} for ATLASGAL candidate starless clumps and \citet{urquhart2013atlasgal} for ATLASGAL sources associated with 6.7~GHz methanol masers. This is likely to be due to the hierarchical structure of molecular clouds from size scales of clouds to cores.

\subsection{Dust temperature}
\begin{figure}
\centering
\includegraphics[width=0.45\textwidth, trim = 0 0.4cm 0 0]{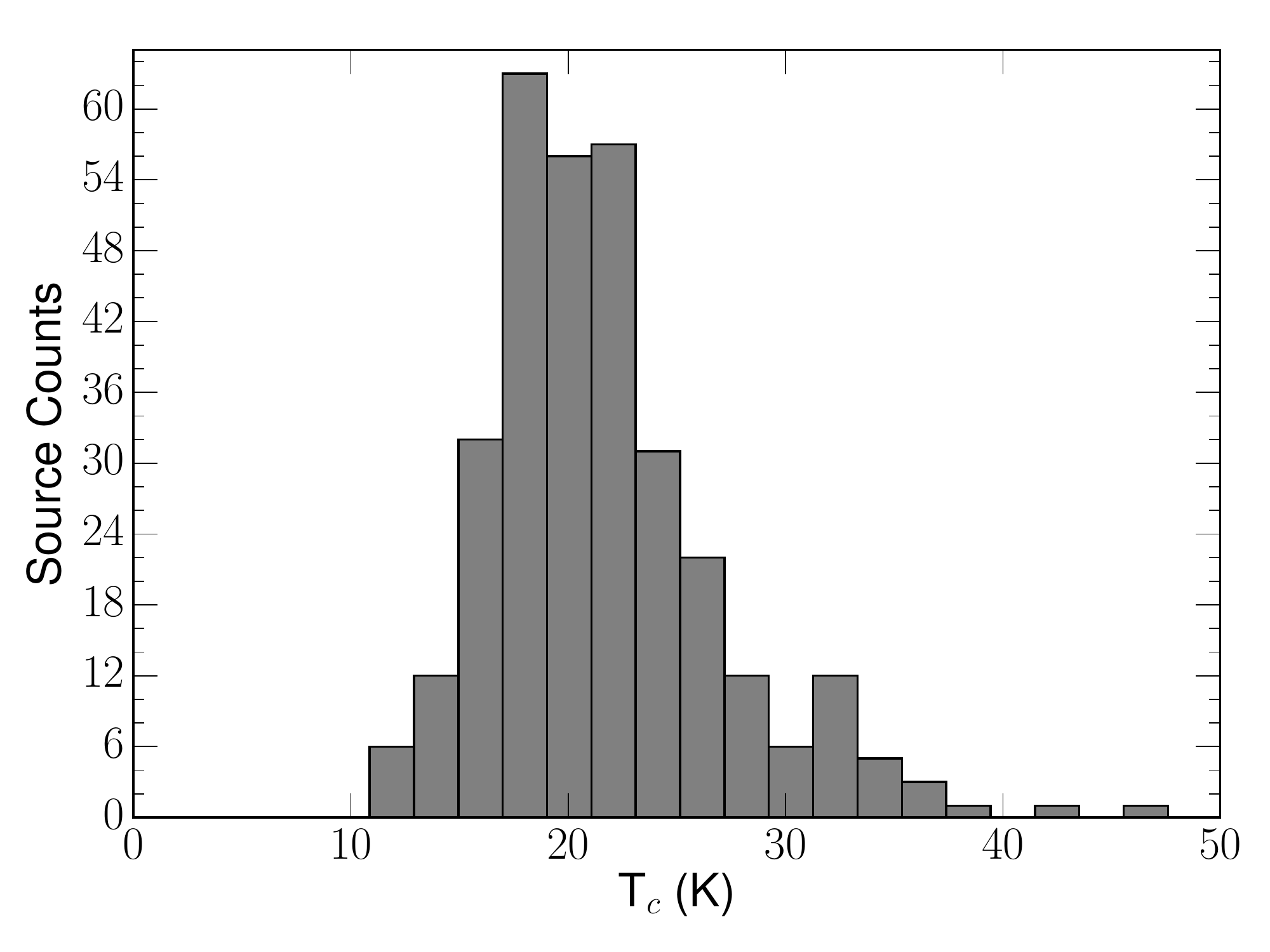}
\caption{Distribution of cold dust  temperature.}\label{fig:tempdistribution}
\end{figure}

The cold dust temperature is found to range from 10.9~K to 47.6~K, with mean and median values of 21.5~K and 20.9~K respectively. Figure \ref{fig:tempdistribution} shows the distribution of dust temperature. The dust temperatures compare well with kinetic temperatures determined from ammonia measurements where a median temperature of 23.4~K was observed \citep{pandian2012physical}. This further corroborates the more evolved nature of 6.7~GHz methanol masers compared to infrared dark clouds where lower temperatures are measured (e.g. \citet{pillai2006ammonia}).

\subsection{Clump Masses, H$_2$ column densities and surface densities} 
\begin{figure*}
\centering
\includegraphics[width=0.45\textwidth, trim= 0 0.4cm 0 0]{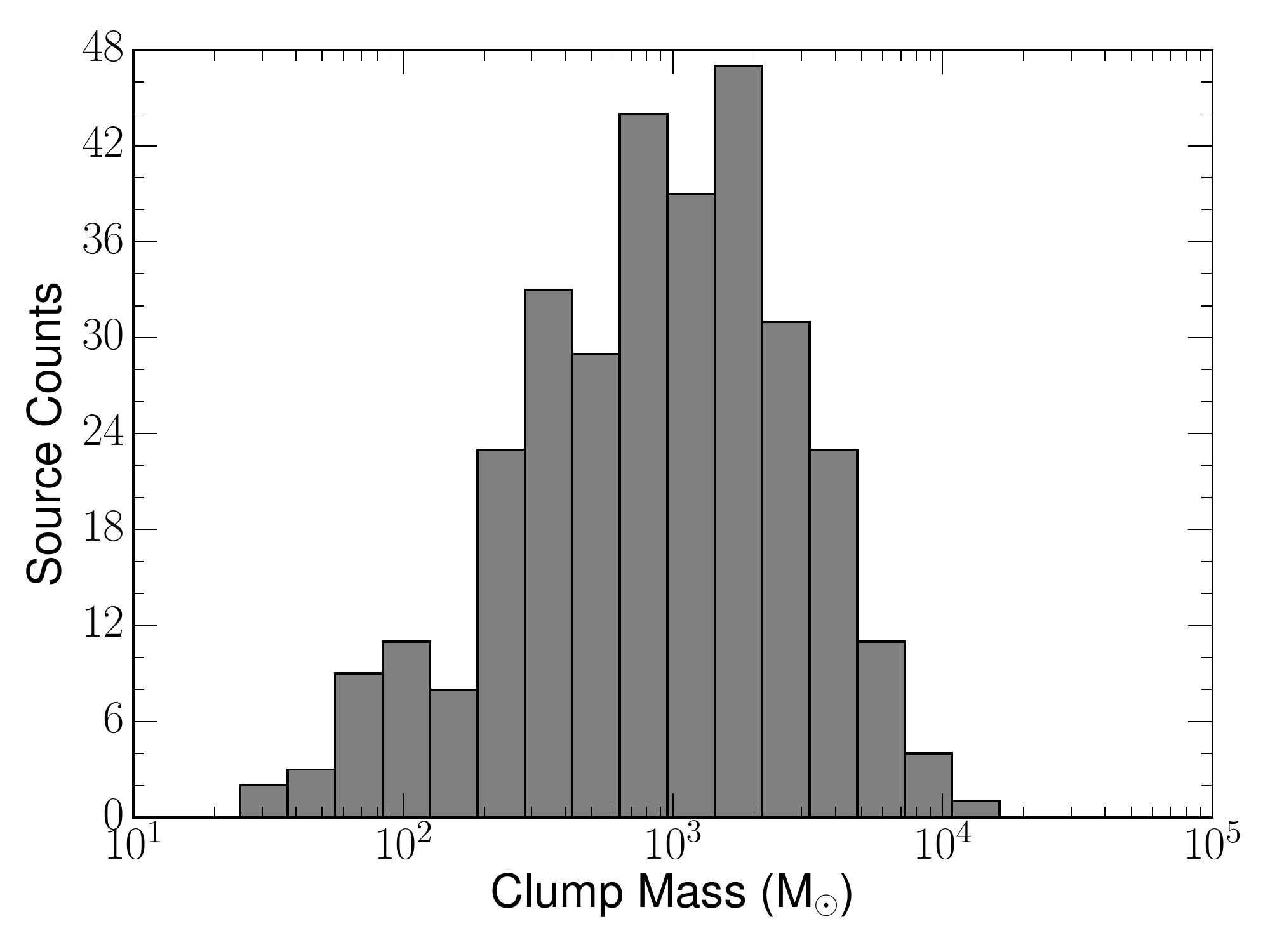}
\includegraphics[width=0.45\textwidth, trim= 0 0.4cm 0 0]{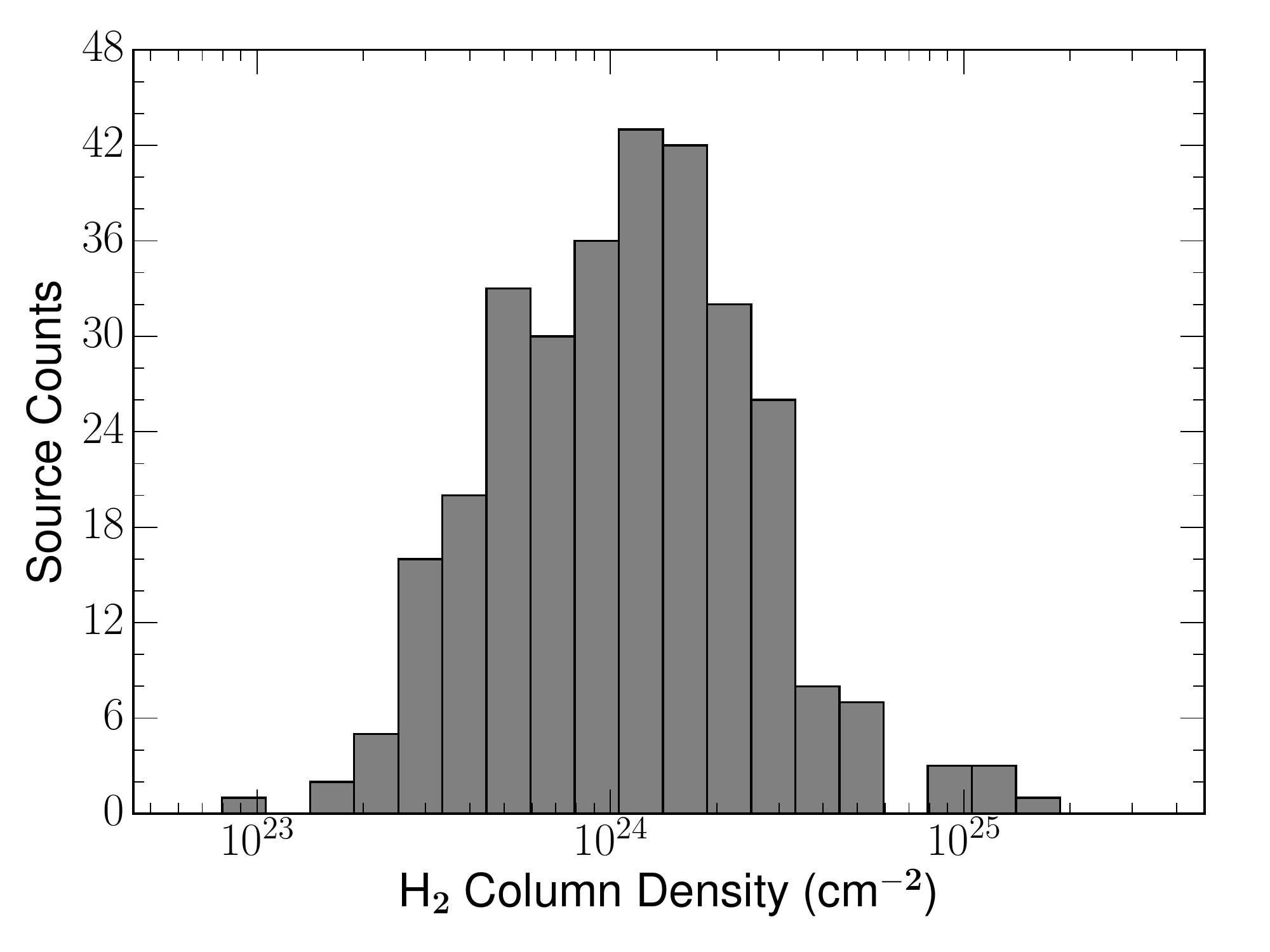}
\caption{Distribution of mass (left panel) and H$_2$ column density (right panel).}\label{fig:massdistribution}
\end{figure*}

The isothermal masses of the sources can be computed using the temperature values obtained from the best fit using the following equation:
\begin{equation}
M=\frac{D^{2}S_{\nu}R}{B_{\nu}(T_{D})\kappa_{\nu}}
\end{equation}
where S$_{\nu}$ is the integrated 870~$\mu$m flux, D is the distance to the source, R is the gas-to-dust mass ratio (assumed to be 100), $B_{\nu}$ is the Planck function for the cold dust temperature $T_{D}$ and $\kappa_{\nu}$ is the dust opacity which is taken as 1.85~cm$^{2}$~g$^{-1}$ at  870~$\mu$m. The clump masses range from 11.07~$M_\odot$ to $1.62 \times 10^{4}$~$M_\odot$, the median mass being approximately 930~$M_\odot$. The mass distribution of the clumps is shown in Fig.~\ref{fig:massdistribution} (left panel). 

The shape of the mass distribution is similar to that observed by \citet{urquhart2013atlasgal} although there are some key differences. While the peak of the distribution is similar, the maximum mass is well below that of \citet{urquhart2013atlasgal}. We also see a larger fraction of sources in the 100 to 1000~$M_\odot$ range. These differences are most likely due two reasons. First, the methodology of doing photometry is different -- the \textsc{\large{hyper}} software uses a Gaussian aperture to do photometry, while the fluxes of \citet{urquhart2013atlasgal} are from the ATLASGAL compact source catalog where photometry has been done by \textsc{\large{sextractor}} \citep{contreras2013atlasgal}. The former measures the flux of the compact source while the latter measures the flux of the entire clump including the diffuse emission around the compact source. To corroborate this, we have compared the fluxes determined by us with those of the GAUSSCLUMP catalog \citep{csengeri2014atlasgal}. We find our fluxes to be comparable though slightly larger than that of \citet{csengeri2014atlasgal}. This is because the latter fit the entire clump including the outer diffuse structure with multiple Gaussians, while we only fit the central emission based on its morphology at 250~$\mu$m which is our reference wavelength. The larger fluxes in the ATLASGAL compact source catalog translate to larger masses in the work of \citet{urquhart2013atlasgal}.

An additional factor that contributes to the different mass distribution is the assumption of a uniform dust temperature of 20~K by \citet{urquhart2013atlasgal} as opposed to our deriving the dust temperature by fitting the spectral energy distribution. As mentioned in the section 3.2, the dust temperature ranges from 10.9~K to 47.6~K, with 185 out of 320 sources having temperatures greater than 20~K. As is evident from eq.(5), assumption of a dust temperature of 20~K leads to the mass being overestimated when the true temperature is greater than 20~K. Since a majority of sources have temperatures greater than 20~K, our masses will be lower than that derived by \citet{urquhart2013atlasgal} even without accounting for the difference in photometry. However, since the mean temperature is very close to 20~K, we see the peak in our mass distribution to be comparable to that of \citet{urquhart2013atlasgal}.
\begin{figure}
\centering
\includegraphics[width=0.45\textwidth, trim= 0 0.4cm 0 0]{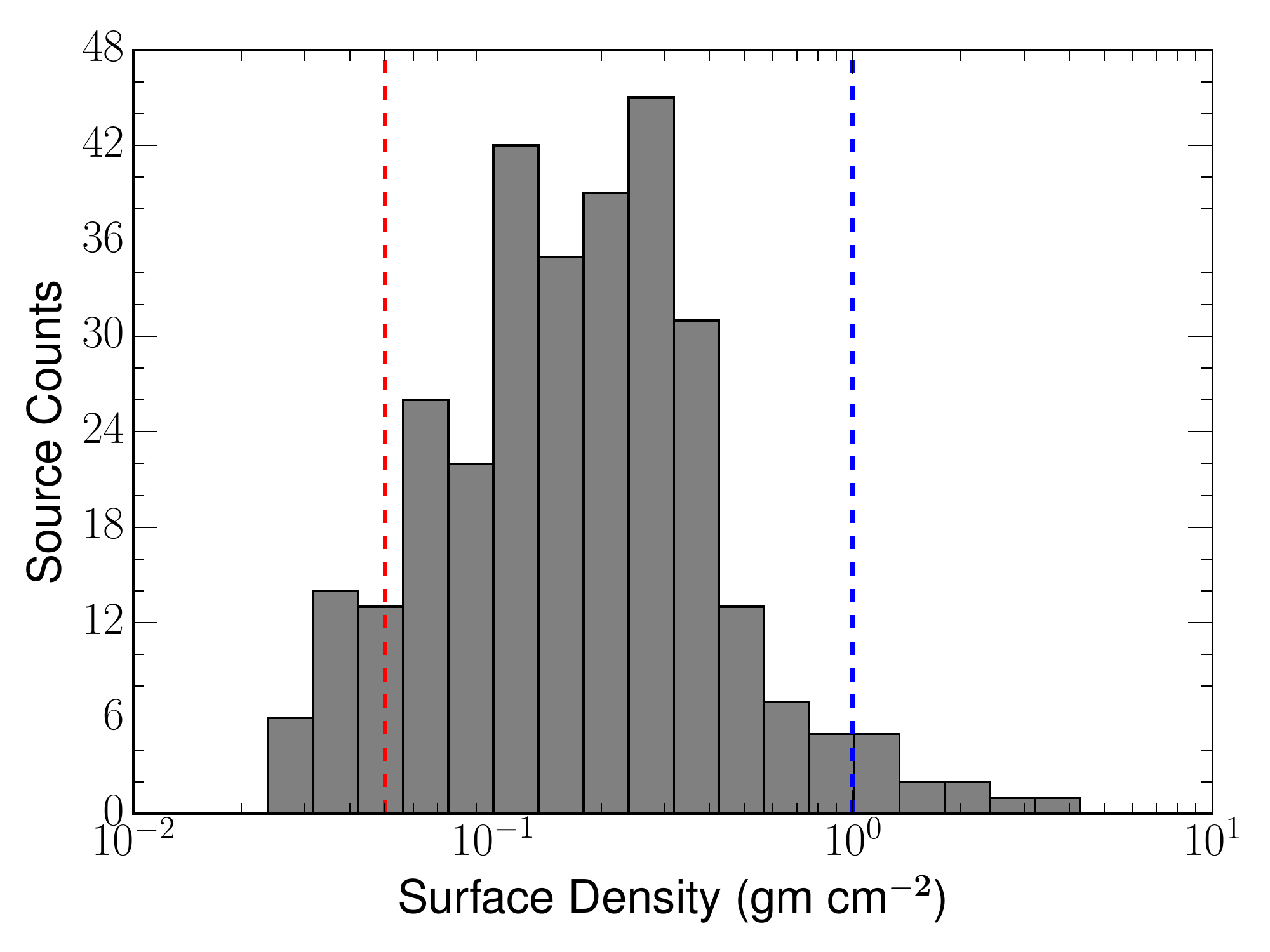}
\caption{Distribution of the surface density. The \textit{red} and \textit{blue} dashed line show a threshold of 0.05~g~cm$^{-2}$ and 1.00~g~cm$^{-2}$ respectively. }\label{fig:sigmadistribution}
\end{figure}

The molecular hydrogen column densities are estimated from the 870~$\mu$m flux density using the following equation: 
\begin{equation}
N_{H_{2}}=\frac{S_{\nu}R}{B_{\nu}(T_{D})\Omega\kappa_{\nu}\mu m_{H}}
\end{equation}
where $\Omega$ is the deconvolved solid angle of the source, $\mu$ is the mean molecular weight and is assumed to be equal to 2.8 (the hydrogen mass fraction is assumed to be $\sim 0.7$) and $m_{H}$ is the mass of an hydrogen atom. It has to be noted that we have determined the column densities of only those sources that have sizes larger than the beam size (since we have used the deconvolved solid angle in eq.(6)). The column densities of the maser associated clumps range from $7.95 \times 10^{22}$ to $ 2.5 \times 10^{25}$~cm$^{-2}$ with a median value of $1.15 \times 10^{24}$~cm$^{-2}$.  The observed values are generally higher than those found in previous studies of high-mass star forming sources (e.g. \citealt{garay2004discovery}) due to the usage of deconvolved rather than observed solid angle of the source. The distribution of column densities of the sample are shown in Fig.~\ref{fig:massdistribution}.

Another parameter of high interest in star formation is the surface density ($\Sigma)$ of the clump or core. The surface densities are obtained by dividing the mass of the clump by its physical area ($\pi R_\mathrm{eff}^2$ where $R_\mathrm{eff}$ is the effective radius). The surface densities inferred for our sample (that are properly deconvolved) range from 0.02 to 4.28~g~cm$^{-2}$ with a mean and median value of 0.27 and 0.17~g~cm$^{-2}$ respectively. Fig.~\ref{fig:sigmadistribution} shows the distribution of surface density for our sample. Also shown in Fig.~\ref{fig:sigmadistribution} is the threshold of 0.05~g~cm$^{-2}$ that is suggested by \citet{urquhart2013atlasgal} as the minimum surface density required for forming massive stars.

\subsection{Clump and maser luminosities}

\begin{figure}
\centering
\includegraphics[width=0.45\textwidth, trim= 0 0.4cm 0 0]{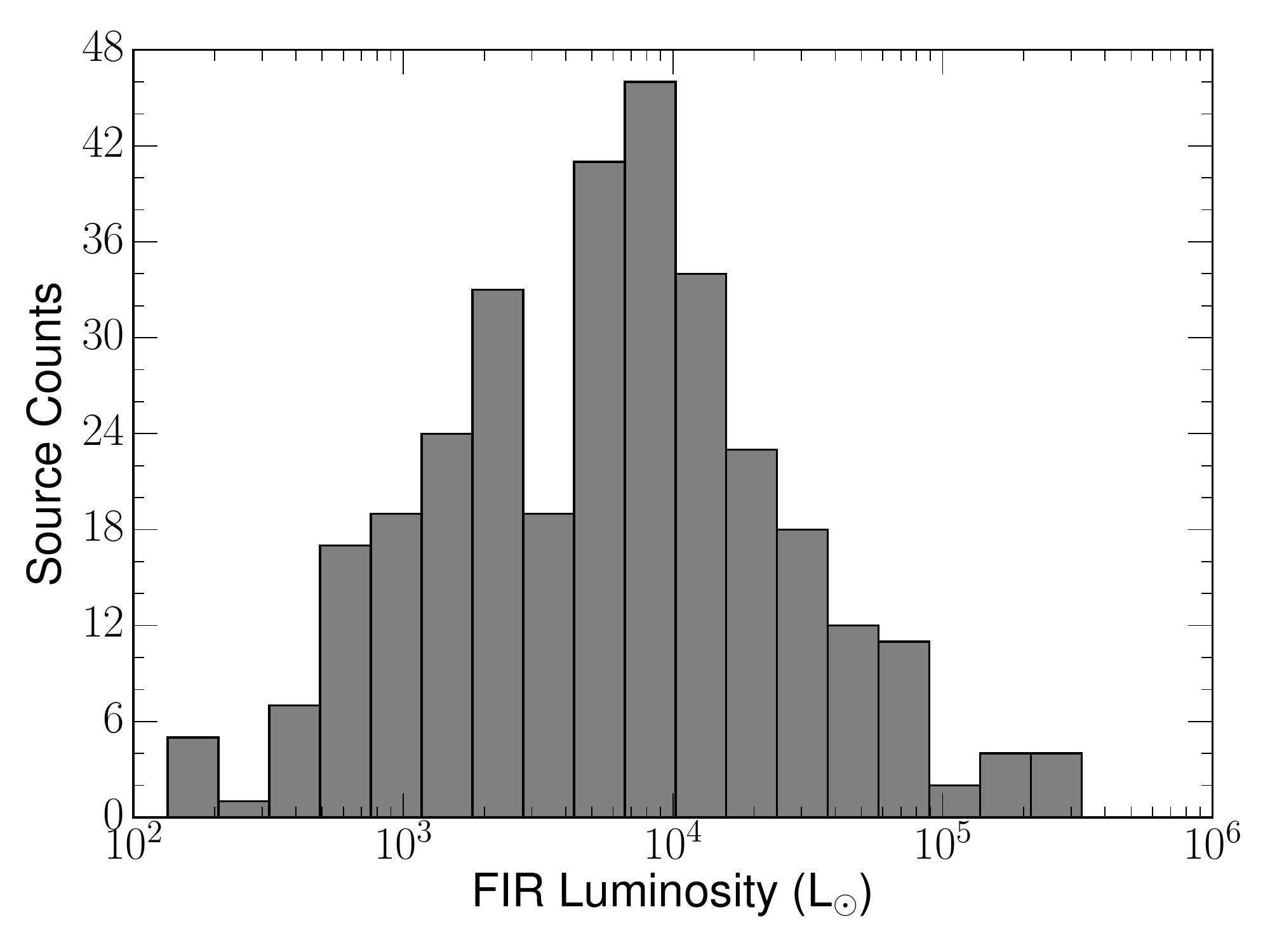}
\caption{Distribution of the far infrared luminosity.}\label{fig:lumdistribution} 
\end{figure}

The luminosity of the individual clumps have been calculated by integrating the SED:
\begin{equation}
L=4\pi D^{2}\int f_{\nu}d\nu
\end{equation}
where D is the distance to the source and $\int f_{\nu}d\nu$ is the integrated flux. We have calculated the Far Infrared (FIR) luminosity using the modified blackbody fit of the cold dust. The FIR luminosity estimates range from $133$~L$_\odot$ to $3.3 \times 10^{5}$~L$_{\text{\ensuremath{\odot}}}$ with mean and median values of $1.74 \times 10^{4}$~L$_\odot$ and $6.4 \times 10^3$~L$_\odot$ respectively. Fig.~\ref{fig:lumdistribution} shows the distribution of the FIR luminosities of the sample.

Although the total IR luminosity is expected to be higher than the FIR luminosity, especially since the methanol maser sources are sufficiently evolved to have emission at 24~$\mu$m and shorter wavelengths covered by the GLIMPSE survey, it is interesting that we detect relatively low luminosity sources with luminosities lower than $10^3$~L$_\odot$ hosting methanol masers. This suggests that there is a small population of methanol masers that are associated with intermediate-mass or low-mass stars.

One of the aims of this paper is to explore the relation between methanol maser luminosity and the global properties of the maser hosts. We have hence computed the isotropic methanol maser luminosities using the following equation:
\begin{equation}
L_{MMB}=4\pi D^{2}S_{\nu}
\end{equation}
where $S_{\nu}$ is the integrated flux across the maser line and is taken from \citet{breen20156}. The methanol maser luminosity values range from $3.31 \times 10^{-10}$ to $1.81 \times 10^{-5}$~L$_\odot$ with mean and median values of $5.05 \times 10^{-7}$~L$_\odot$ and $6.3 \times 10^{-8}$~L$_\odot$ respectively.

The left panel of Fig.~\ref{fig:lumdistribution} shows the clump mass as a function of the maser luminosity. A partial Spearman correlation test to remove the mutual dependence on the distance to the source gives a correlation coefficient of 0.27. The right panel of Fig.~\ref{fig:lumdistribution} shows the FIR luminosity of the clumps as a function of the maser luminosity, the partial Spearman coefficient between these being 0.37. These results are similar to that observed by \citet{urquhart2014atlasgal} wherein a weak correlation was seen between the clump mass and maser luminosity. It is to be noted that \citet{urquhart2014atlasgal} computed the maser luminosity using the peak line flux (i.e. in Jy~kpc$^2$) rather than integrated line flux. However, we verify that the values of correlation coefficients obtained using methanol maser luminosities computed from their peak line fluxes are very similar to those reported above using the integrated line flux.

One of the concerns with the work presented here is the limited resolution of the data, which when coupled with the large distances to the sources could result in more than one compact source (which may be in a different evolutionary state) being observed within the same telescope beam. However, as indicated in section 2, a large fraction of sources ($\sim 90\%$) are not observed to be resolved into multiple sources at the 10$''$ resolution of the Hi-GAL 70~$\mu$m data. An additional way to test whether multiplicity affects the results presented here is to divide the sample into two groups, based on a distance threshold. We computed the correlation coefficient based on the partial Spearman test for the two groups for different values of the distance threshold. We found that as long as the masers with maser luminosity greater than $10^{-6}$~L$_\odot$ were excluded from the sample, both groups had similar correlation coefficients that were less than 0.2 (irrespective of the value used for the distance threshold) which is consistent with no correlation. However, the correlation coefficient increased significantly when including the masers with maser luminosity greater than $10^{-6}$~L$_\odot$. This correlation was found to be driven by the relatively high FIR luminosity of the clumps hosting the masers with high maser luminosity. Since the number of such sources is relatively small, it is not possible to draw further conclusions on whether this is a systematic effect. However, if verified, this suggests that a high FIR luminosity is required to pump methanol masers with high maser luminosity, which is consistent with the radiative pumping mechanism for the masers \citep{sobolev2007methanol}. The weak to non-existent correlation between the FIR luminosity and maser luminosity for the rest of the sample suggests that other factors such as the density, methanol fractional abundance and the gas kinetic temperature in the masing spots are more important to determining the maser luminosity compared to the density of pumping photons.

\begin{figure*}
\begin{center}
\includegraphics[width=0.45\textwidth, trim= 0 0.4cm 0 0]{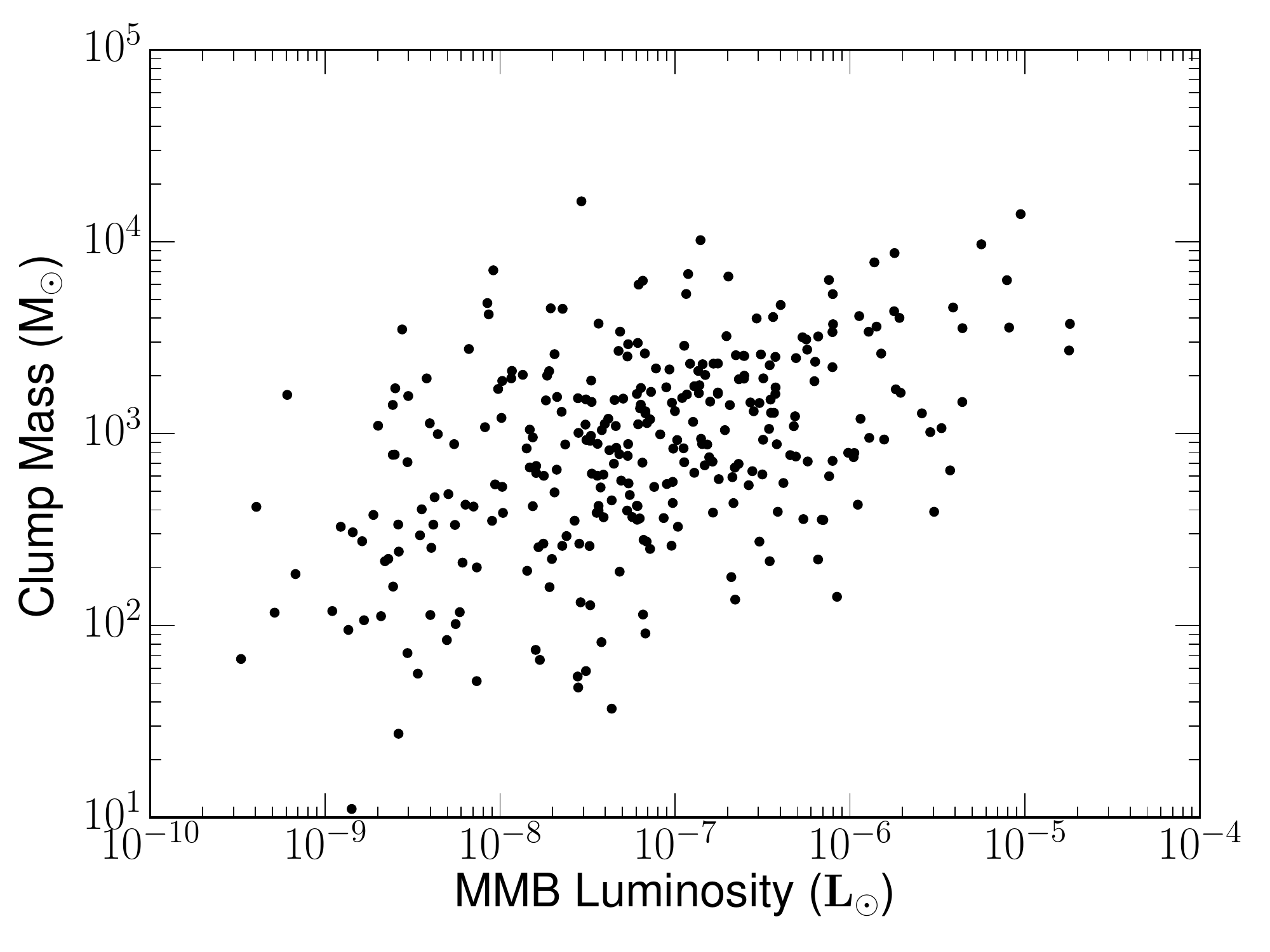}
\includegraphics[width=0.45\textwidth, trim= 0 0.4cm 0 0]{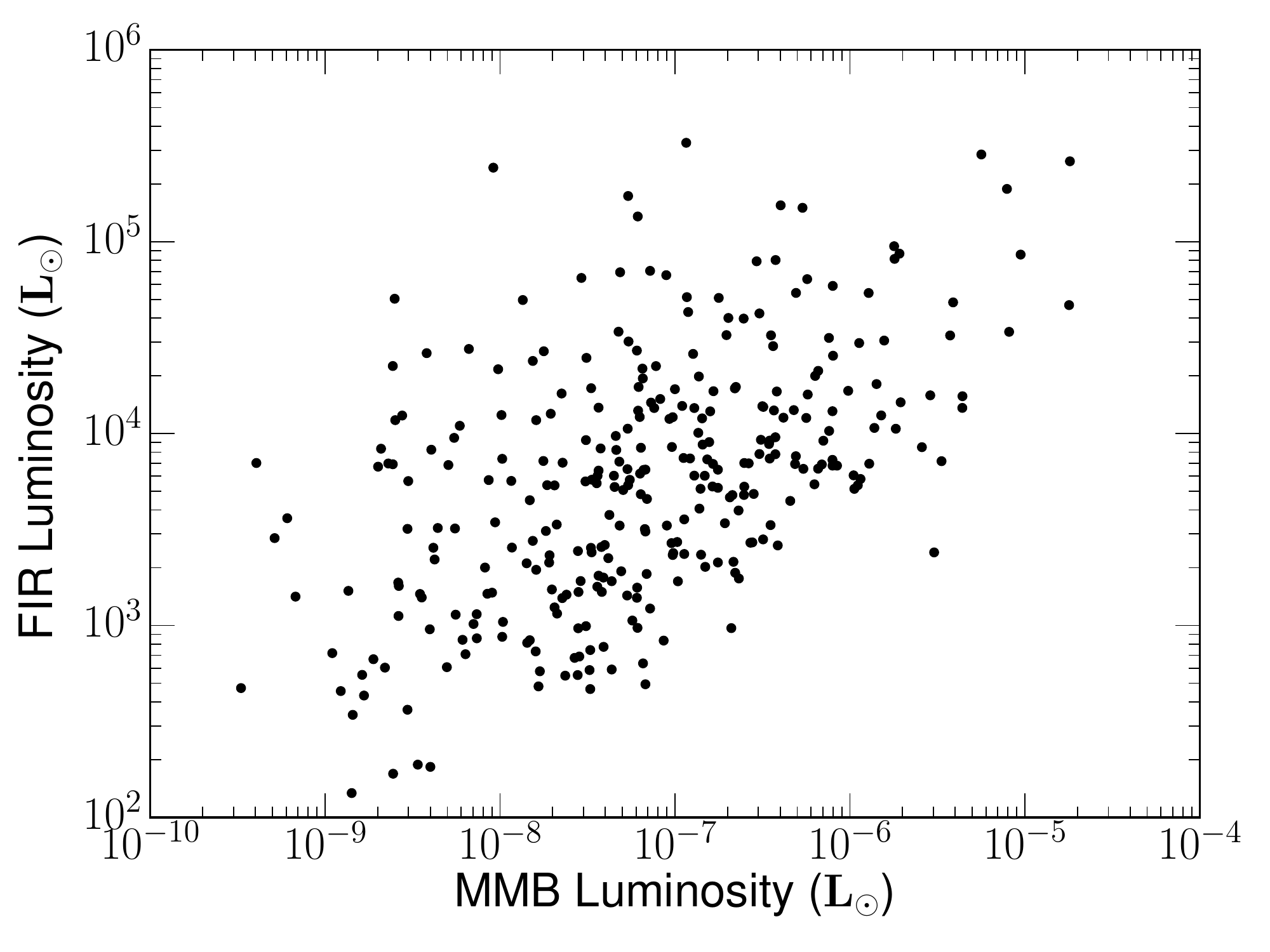}
\caption{Relationship of 6.7 GHz methanol maser luminosity with (a) clump mass (left panel) and (b) FIR luminosity (right panel). }\label{fig:MMrel} 
\end{center}
\end{figure*}

\begin{figure}
\centering
 \includegraphics[width=0.45\textwidth, trim= 2cm 0.4cm 0 0]{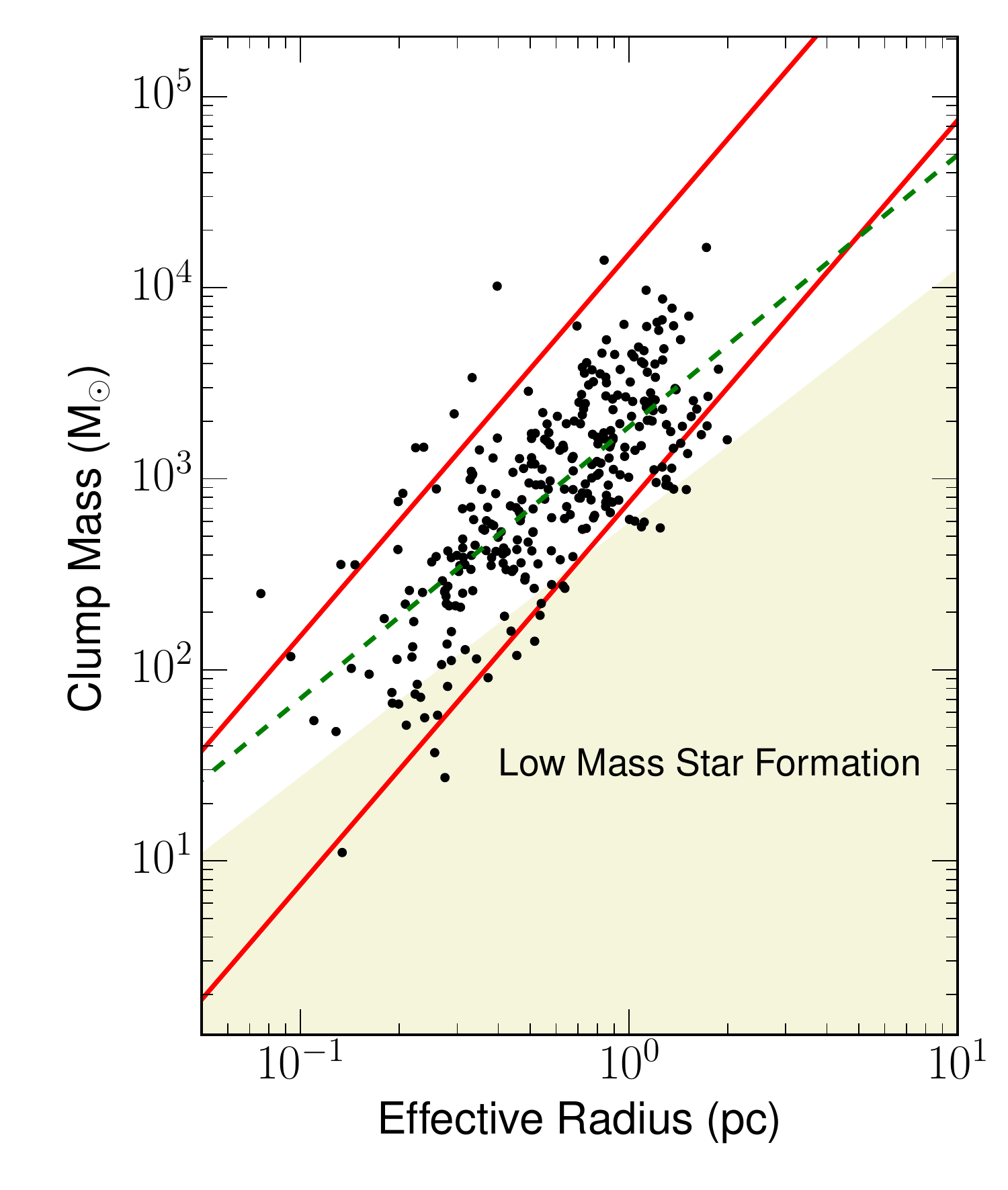}
\caption{The clump mass as a function of source radius. The shaded region represents the area where the sources doesn't satisfy the Kauffman~criteria. The \textit{dashed green} line is the power law fit to the data. The upper and lower \textit{solid red line} shows the surface densities of 1~g~cm$^{-2}$ and 0.05~g~cm$^{-2}$, respectively. }
\label{fig:MRrelation}
\end{figure}

\begin{figure*}
\begin{center}
\includegraphics[width=0.45\textwidth, trim= 0 0.4cm 0 0]{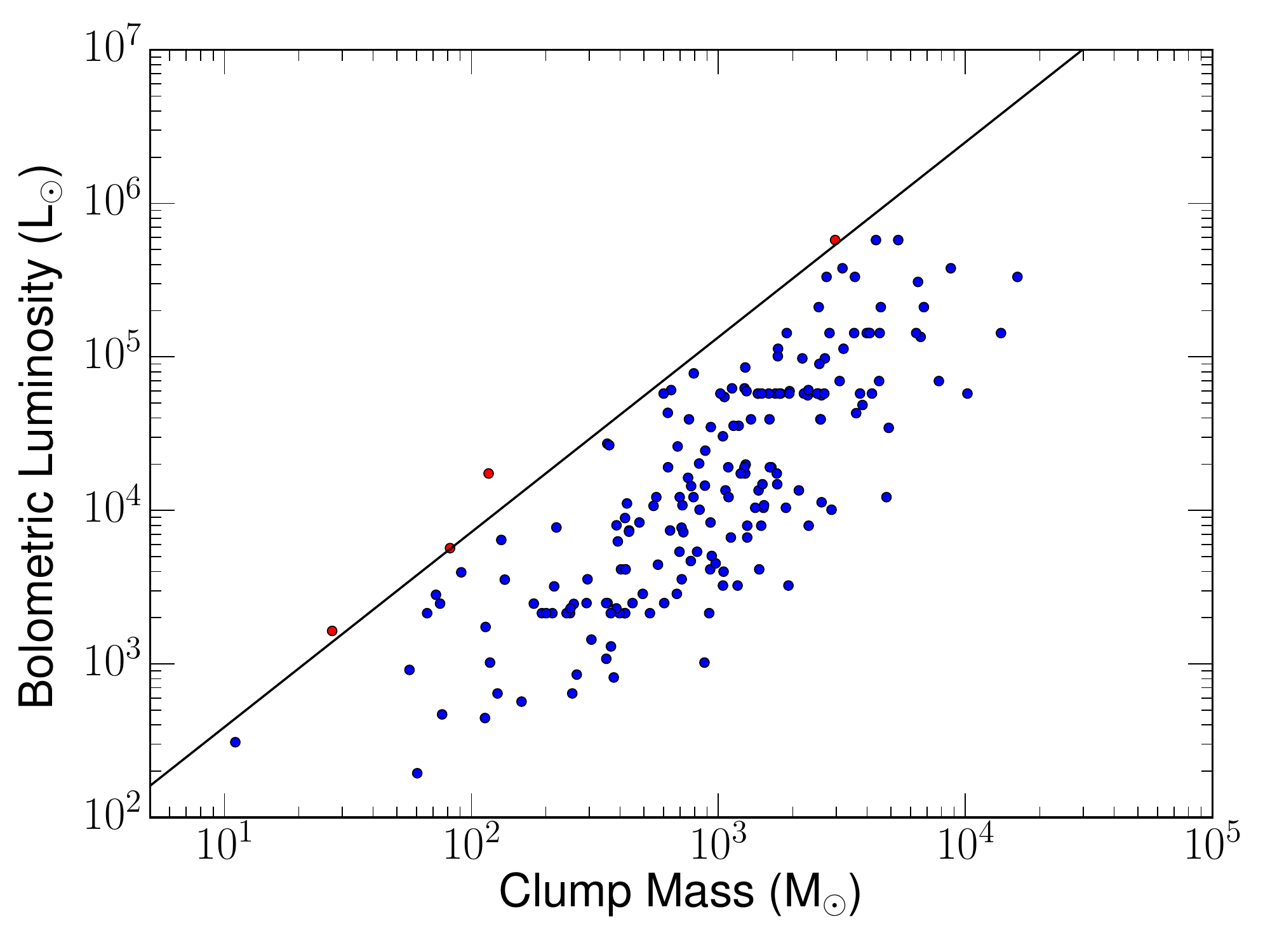}
\includegraphics[width=0.45\textwidth, trim= 0 0.4cm 0 0]{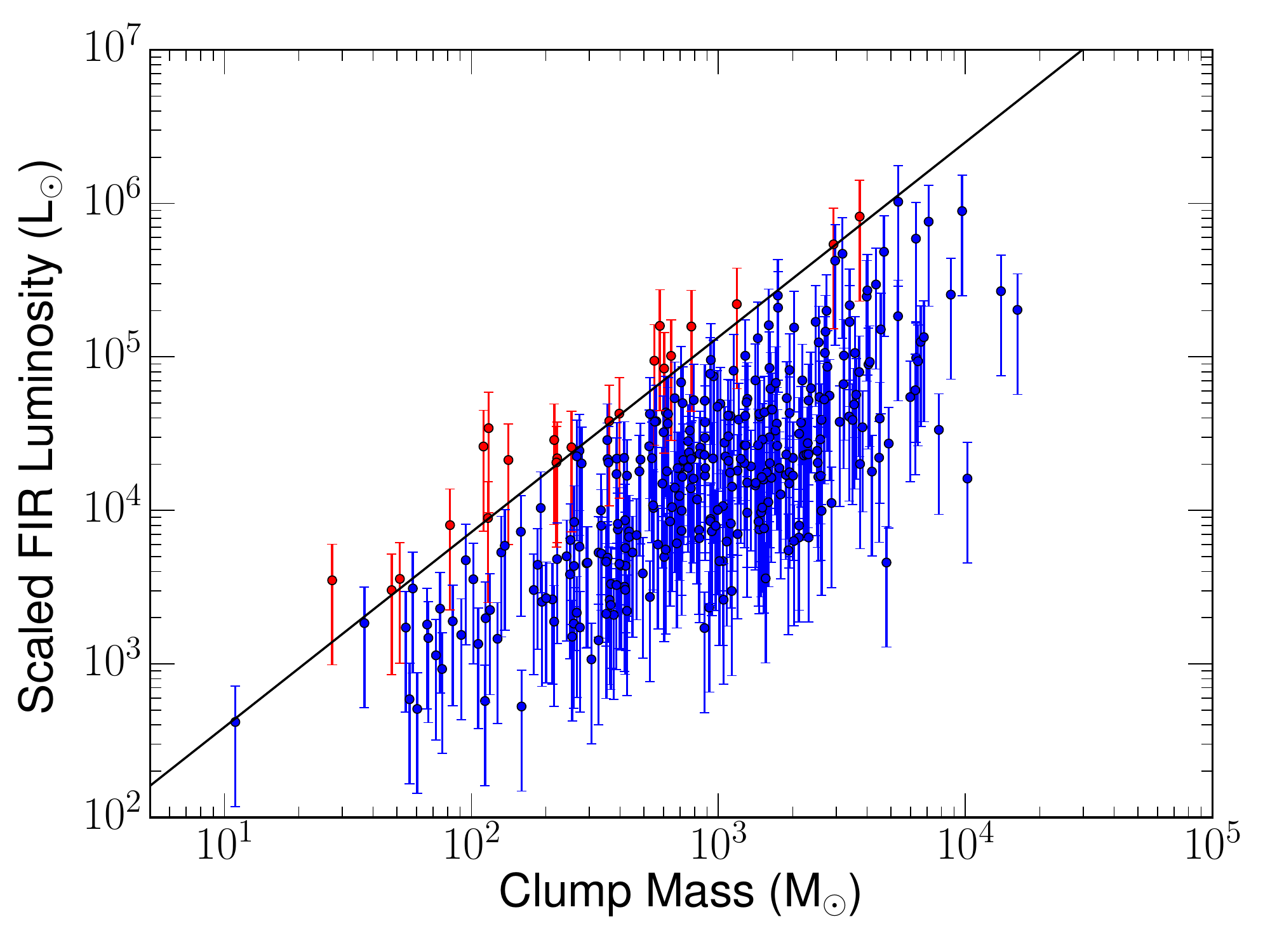}
\caption{The left panel shows the bolometric luminosity of 198 sources that has MIPSGAL counterparts as a function of mass.The right panel illustrates the scaled FIR luminosities as a function of clump mass. The solid line represents the fit to the ``IR-P'' sources in \citet{molinari2008evolution}. The methanol maser hosts that are in accretion phase and clearing phase are shown in \textit{blue} and \textit{red} points respectively.} \label{fig:MLrel}
\end{center}
\end{figure*}

\section{DISCUSSION}\label{sec4}

In this section, we address the questions of whether or not 6.7~GHz methanol masers are exclusively associated with massive star formation, and the evolutionary phase of the young stellar objects that are associated with the masers.

At the outset, we remind the reader that the majority of sources in our sample are at large distances and almost all the clumps that are seen as isolated structures at 18$''$ resolution are likely to fragment into smaller structures at high spatial resolution. The clump properties derived from single dish observations will be representative of the most massive core within the clump. Thus, discussion of the clump properties in the context of their hosting a methanol maser makes the implicit assumption that the maser is excited by the most massive core, which is not necessarily true. However, \citet{Chibueze_2017} has explored the association of 6.7 GHz methanol masers with massive dense cores (MDC) detected with the Atacama Compact Array (ACA) at 4$''$ resolution, and find a direct association between the MDC and the methanol maser in 91\% of their sample. A targeted 1.05 mm ALMA observtion towards the deeply embedded high-mass protocluster G11.92$-$0.61 by \citet{Cyganowski_2017} also reveals the association of 6.7 GHz methanol masers with the massive cores in this region. These studies suggest that the 6.7 GHz methanol maser emission is indeed mostly associated with the most massive core in the clump. However, a caveat must be borne in mind that there may be a small number of cases where the maser emission may arise from a source other than the most massive core in the clump.

\subsection{Mass-radius relation}
The mass-radius relationship of nearby molecular cloud complexes was investigated by \citet{kauffmann2010massa,kauffmann2010massb} who then compared the relation with those of known high-mass star forming regions. This led to a suggestion that clumps with the potential to form at least one massive star follow the following empirical relationship:
\begin{equation} 
m(r) \geq 580~M_\odot\left(\frac{R_\textrm{eff}}{1~\textrm{pc}}\right)^{1.33}
\label{kauffmanneqn}
\end{equation}
where $R_\mathrm{eff}$ is the effective radius of the source. Fig.~\ref{fig:MRrelation} shows the mass-radius relationship for our sample. We find that 295 out of 320 sources (92\%) satisfy the criterion in the equation above. The results are somewhat different from that of \citet{urquhart2013atlasgal,urquhart2014almost} who found a much larger fraction ($\sim 97\%$) of the methanol masers satisfying the criterion above. This discrepancy is most likely to be due to two factors. First, the photometry is different as explained in section 3.2. The compact source catalog used by \citet{urquhart2013atlasgal} is sensitive to the entire submillimeter clump including the diffuse emission, while the photometry using \textsc{\large{hyper}} is sensitive only to the compact emission, similar to the catalog of \citet{csengeri2014atlasgal}. A second factor is that we have determined effective radii using the 250~$\mu$m Hi-GAL maps (which we have used as the reference wavelength for \textsc{\large{hyper}}) rather than the 870~$\mu$m ATLASGAL maps. The former typically has much better signal to noise ratio leading to a better estimation of the source size. To further examine this matter, we looked at the sources in our sample that do not satisfy the Kauffmann criterion and compared the effective radius given in the compact source catalog with that of our work. We found that the effective radius is listed for only 5 out of 25 sources, the remaining being too compact for deconvolution. We found the effective radius and the enclosed flux to be larger in the compact source catalog compared to the \textsc{\large{hyper}} results leading to the former satisfying the Kauffmann criterion. Although the number of sources examined here is small, it seems to confirm the hypothesis of the difference between our results and those of \citet{urquhart2013atlasgal} as primarily arising from the differences in photometry.

Fig.~\ref{fig:MRrelation} also shows that the mass of the clump is correlated to its effective radius. A partial Spearman correlation test was done to remove any dependence of the correlation on distance yielding a correlation coefficient of 0.22, showing that the two quantities are weakly correlated. A least squares fit to the data gives the mass-radius relationship for our sample as 
\begin{equation}
\log M_\textrm{clump} = 3.27 + 1.42 \log R_\textrm{eff}
\end{equation}
This is somewhat different from the results of \citet{urquhart2013atlasgal} who found a power law index of 1.67. The upper diagonal line in Fig.~\ref{fig:MRrelation} shows a line with a constant surface density of 1~g~cm$^{-2}$, which is suggested to be the threshold for the formation of massive stars from turbulent cores \citep{mckee2003formation}. It can be seen that only a small fraction of the sources satisfy the threshold above. However, it must be borne in mind that the threshold of \citet{mckee2003formation} applies to cores which will form one or two massive stars. As explained in section~3.1, the resolution of the Hi-GAL and ATLASGAL data is such that most of the structures seen in our images are clumps rather than cores. Using the mass-radius relation, \citet{urquhart2013atlasgal} suggested that a surface density of 0.05~g~cm$^{-2}$ (lower red line in Fig.~\ref{fig:MRrelation}) provided an empirical lower bound for the clump surface density required for massive star formation, although it differs from the Kauffmann criterion for low values of effective radius. A total of 293 sources in our sample have surface densities higher than 0.05~g~cm$^{-2}$ agreeing with the results of \citet{urquhart2013atlasgal}.

\subsection{Evolutionary stage of the source}

The evolutionary state of the source can be inferred from a plot of the source luminosity as a function of mass. For a given clump mass, the luminosity is expected to increase as star formation progresses in the clump. Thus, larger values of the L/M ratio are indicative of more evolved sources. In order to construct a L-M diagram, one needs to compute the bolometric luminosity (L$_\mathrm{bol}$) of the sources. Since 6.7~GHz methanol masers are pumped by warm dust, the maser hosts have significant emission at mid-infrared wavelengths, indicated by the presence of counterparts in the MIPSGAL survey at 24~$\mu$m and at shorter wavelengths in the GLIMPSE survey. Hence, the FIR luminosity estimated from the ATLASGAL and Hi-GAL data will be an underestimate of the bolometric luminosity.To address this issue, we constructed the full SED to near-infrared wavelengths for 198 sources using the fluxes in the MIPSGAL and GLIMPSE catalogues (the remaining sources had no counterpart in the MIPSGAL catalogue, presumably due to saturation effects). We then fit the SED from 870~$\mu$m to 3.6~$\mu$m using the models of \citet{robitaille2007interpreting}. The choice of models is not very important since the purpose of this exercise is not to determine the properties of the embedded young stellar object, but rather to obtain the bolometric luminosity by fitting the full SED from submillimeter to near infrared. We found the FIR to bolometric luminosity ratio to have a median value of 0.31 with a standard deviation of 0.23. 

The left panel Fig.~\ref{fig:MLrel} shows the bolometric luminosity as a function of mass for the 198 sources whose SED was fit using the \citet{robitaille2007interpreting} models. The right panel shows the FIR luminosity that is scaled by the ratio of bolometric to FIR luminosity as a function of mass. A partial Spearman correlation test to remove the dependence on distance gives the correlation coefficient between bolometric luminosity and mass (Fig.~\ref{fig:MLrel}, left panel) to be 0.43. The same correlation test when performed on scaled FIR luminosities and clump mass for the entire sample (Fig.~\ref{fig:MLrel}, right panel) yielded a correlation coefficient of 0.42 with a $p$ value $\ll 0.05$ showing that the bolometric luminosity is weakly correlated to the mass of the clump. The correlation is however smaller than that observed by \citet{urquhart2013atlasgal} who observed a correlation coefficient of 0.78.

\citet{molinari2008evolution} constructed a model for the evolution of a source in the L-M diagram based on the turbulent core model of \citet{mckee2003formation}. This model shows the evolution to be in two stages -- in the initial phase, the central star accretes matter from the envelope with the accretion rate being dependent on the instantaneous stellar mass. Thus, as the stellar mass increases with time, so does the accretion rate leading to this phase being referred to as the accelerating accretion phase. During this phase, the envelope mass is almost constant while the luminosity of the source increases leading to vertical tracks in the L-M diagram (see Fig.~9 of \citealt{molinari2008evolution}). This is followed by the envelope clean-up phase wherein the envelope is expelled through outflows and accretion onto lower mass objects in the same clump. Since the luminosity is dominated by that of the massive star (which has reached its final mass), this phase corresponds to horizontal tracks to the left in the L-M diagram. One of the attractive features of this model is that it is a natural extension of the evolution in the low-mass regime \citep{saraceno1996evolutionary}.

Although we do not have the evolutionary tracks modelled by \citet{molinari2008evolution}, a rough boundary between the accretion phase and envelope clearing phase is obtained by fitting the ``IR-P'' sources (the primary sources in the targeted fields whose SEDs can be fit with a model of an embedded zero age main sequence star) in \citet{molinari2008evolution}. This is shown as a solid line in the left and right panels of Fig.~\ref{fig:MLrel}, with sources below and above this line being color coded as blue and red respectively. Comparing this figure with Fig.~9 of \citet{molinari2008evolution} shows two prominent results: First, almost all sources including the lowest mass source lie in the high-mass regime spanned by the ``IR-P'' sources of \citet{molinari2008evolution}. The other prominent result is that most of the sources ($\sim 93\%$) are in the accretion phase. It has to be noted that in the L$_{bol}$-M plot shown in the left panel of Fig.~\ref{fig:MLrel}, almost all the sources lie in accretion phase. However, this is most likely a selection effect due to SEDs being fit for only those sources that have a counterpart in the MIPSGAL catalogue. The lack of a counterpart in the MIPSGAL catalogue is most likely due to the source being saturated at 24~$\mu$m and thus in a later evolutionary phases.To verify this hypothesis, we constructed SEDs for 20 sources that had no counterpart in the MIPSGAL catalogue using fluxes from the MSX \citep{benjamin2003glimpse} and GLIMPSE \citep{1996AJ....112.2862E} catalogues. We found about 15\% of the sources to be in the clearing phase which is consistent with the overall fraction of $\sim 10\%$ to be in the clearing phase based on scaled FIR luminosities.

The results above however must be treated with caution. First, it is based on the turbulent core model for massive star formation, and alternate theories exist for forming massive stars (e.g. competitive accretion model of \citealt{bonnell2001competitive}). Second, the L-M diagram does not give information about the surface density of a clump which is one of the factors that determine whether or not it will form a massive star. Thus, some sources which fail the \citet{kauffmann2010massa} criterion based on the clump masses and radii are located in the high-mass end of the L-M diagram. However, bearing these caveats in mind, the overall results including statistics are consistent with the findings of \citet{pandian2010spectral} wherein most 6.7~GHz methanol masers are associated with rapidly accreting massive stars, with $\sim 80\%$ being in phases earlier than ultracompact H~\large ii regions (i.e. in the accelerating accretion phase).

\subsection{Are 6.7 GHz methanol masers exclusively associated with massive star formation?}

According to \citet{lada2003embedded} and \citet{motte2003massive}, stellar clusters form from clumps with masses more than 100$-$1000~M$_\odot$ and radii 0.5$-$1~pc. Assuming that the stars formed in clusters follow the initial mass function of \citet{kroupa2001variation}, the total stellar mass in a cluster with at least one 8~M$_\odot$ star is around 110~M$_\odot$. Assuming a star formation efficiency of 30\%, the minimum mass that a clump must have to form a cluster with at least one massive star is 360~M$_\odot$. In our sample, 187 sources have effective radii above 0.5~pc, and 179 out of 187 have masses more than 360~M$_\odot$. Examining sources with effective radii below 0.5~pc, the masses range from 11 to  $1.02 \times 10^{4}$~M$_\odot$. One can thus conclude that most of the 6.7~GHz methanol hosts have masses sufficient to form at least one massive star.

However, there is a small population of 6.7~GHz methanol masers which may be associated with intermediate or low-mass stars. For example, the minimum mass in our sample is 11~M$_\odot$ which is likely to form a star $< 8$~M$_\odot$ depending on the fraction of mass that goes to the central star. A similar conclusion was inferred by \citet{urquhart2013atlasgal} although their masses were obtained by assuming a constant dust temperature of 20~K. The survey \citet{minier2003protostellar} towards 175 low-mass young stellar objects detected 6.7~GHz methanol maser emission towards the source NGC 2024: FIR 4. While the nature of this source has been debated (e.g. \citealt{choi2015radio} and references therein), \citet{choi2015radio} conclude that the source is a low-mass protostar based on analysis of several archival data sets from the Very Large Array. The minimum luminosity of a source that is associated with 6.7~GHz maser emission also appears to be somewhat lower than the $10^3$~L$_\odot$ that is estimated by \citet{bourke2005identification}.

The mechanism by which 6.7~GHz methanol masers are excited by low-mass protostars is however not clear. The dust temperatures required to pump the line are expected to be at distances where the H$_2$ number density is high enough to quench maser action in low-mass protostars \citep{pandian2008detection}. Thus, 6.7~GHz maser action in low-mass protostars may be restricted to select geometries wherein the physical conditions for maser pumping are satisfied. This may also be the reason why the vast majority ($> 95\%$) of the methanol masers are associated with high-mass star formation.\begin{figure}
\centering
 \includegraphics[scale=0.4, trim= 0 0.4cm 0 0]{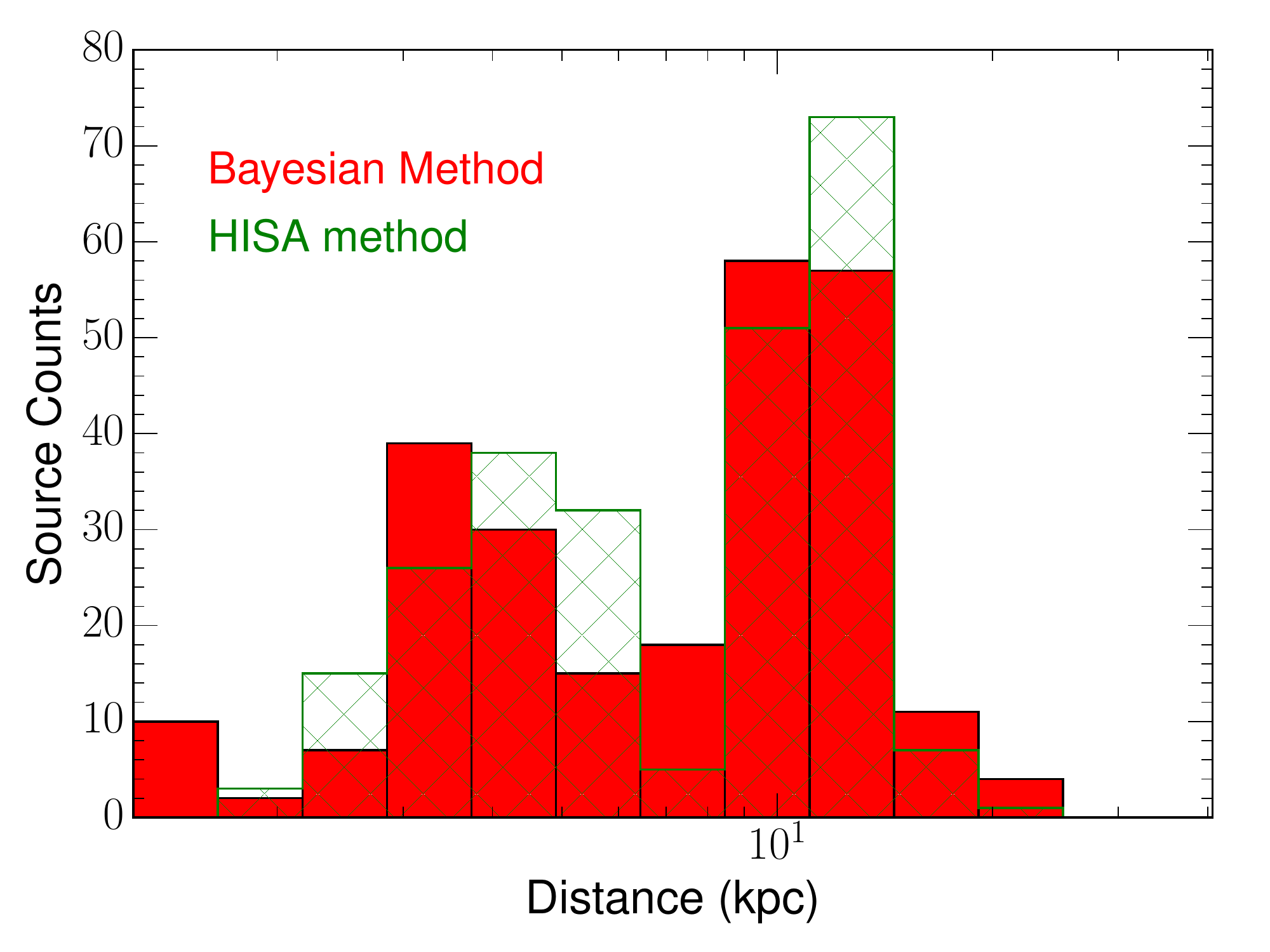}
\caption{Distance distribution based on Method 1 and 2 described in section 4.4}
\label{fig:distcomp}
\end{figure}
\begin{figure*}
\begin{center}
\includegraphics[width=0.45\textwidth, trim= 0 0.4cm 0 0]{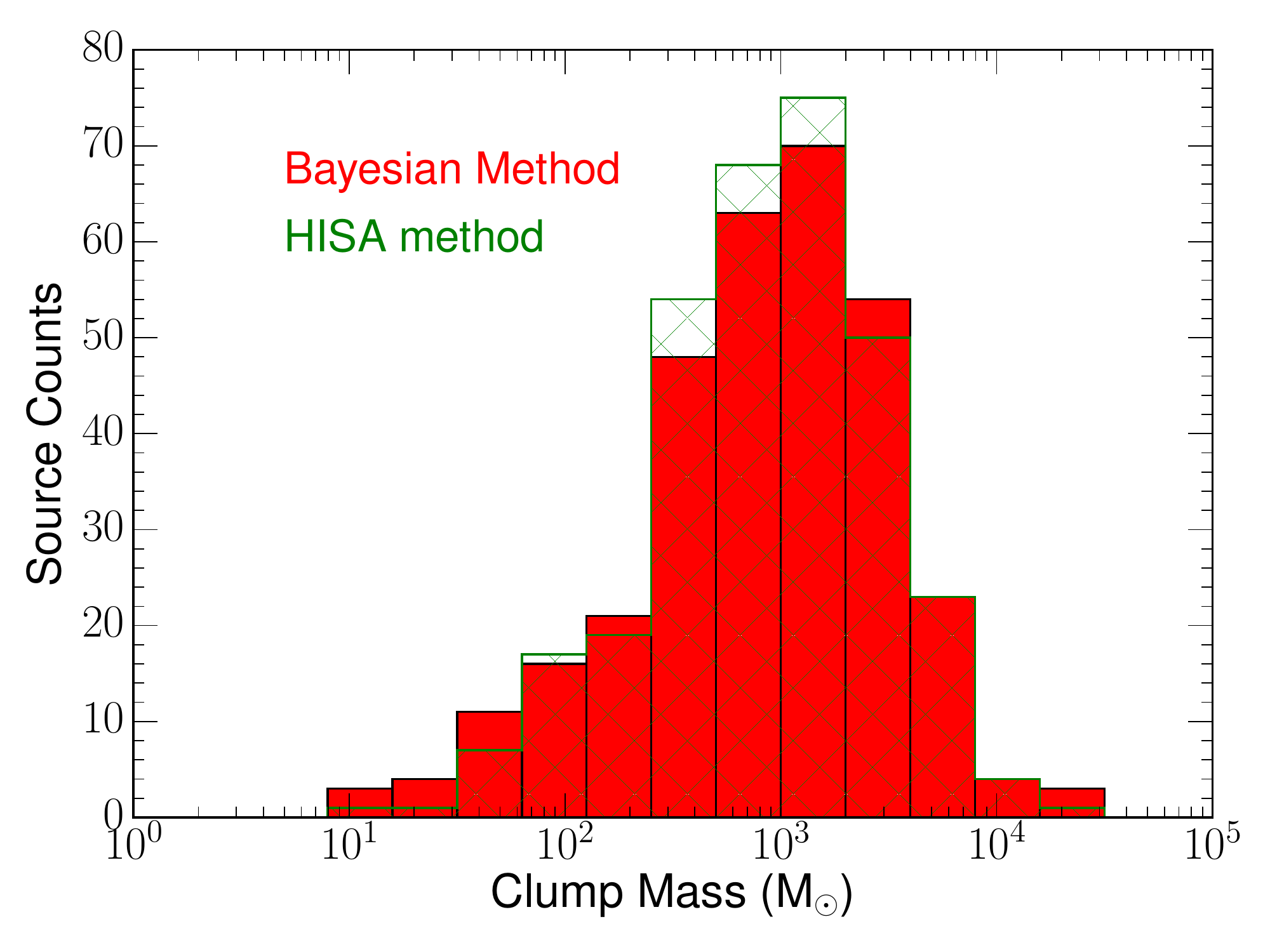}
\includegraphics[width=0.45\textwidth, trim= 0 0.4cm 0 0]{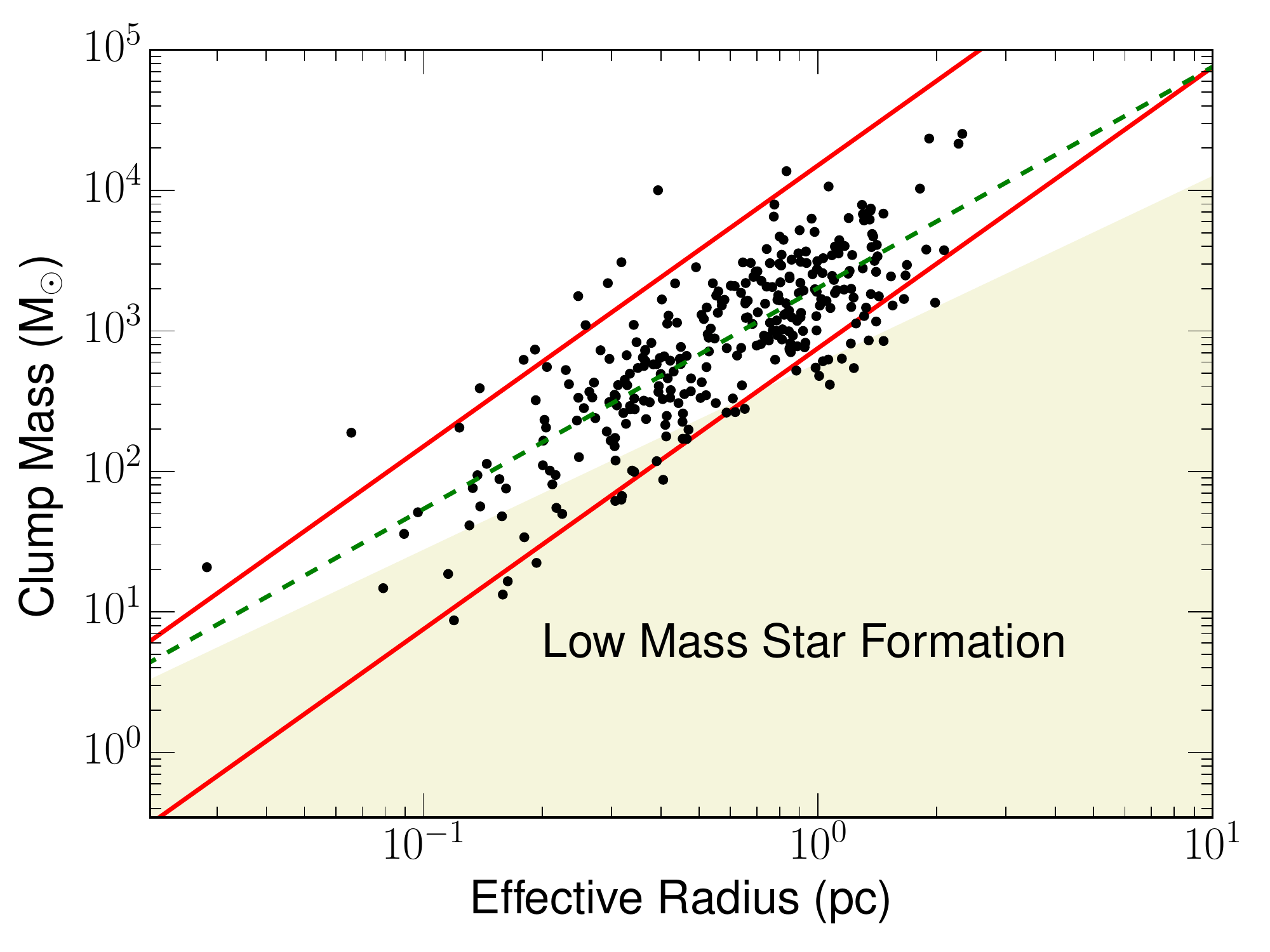}
\caption{Mass distribution using two methods (left panel) and (b) MR plot obtained using the distances obtained with the approach presented by \citet{reid2016parallax}(right panel).The \textit{dashed green line} represents the power law fit to the data. The upper and lower \textit{solid red line} shows the surface densities of 1~g~cm$^{-2}$ and 0.05~g~cm$^{-2}$, respectively.}\label{fig:Mandmr} 
\end{center}
\end{figure*}

\subsection{A special note on distances} 
The most reliable way to estimate the distance to Galactic sources is trigonometric parallax. Due to difficulties in measuring the trigonometric parallax to distant sources, especially in the Galactic disk where extinction is severe, distances are commonly estimated kinematically by assuming that the observed radial velocity is a result of differential rotation of the Milky Way. However, the kinematic distance can be discrepant from the true distance when proper motions, such as from spiral density waves, are significant. The method of calculating distances from Galactic kinematics also suffers from an ambiguity between two distances in the inner Galaxy, requiring use of additional techniques such as {\rm H}\kern 0.1em{\sc i} Self Absorption (HISA) in order to distinguish between the two values.

As mentioned in section~\ref{sec2}, the distances to the sources in our sample are taken from \citet{urquhart2013atlasgala,reid2014trigonometric,pandian2009arecibo} and \citet{green20176}. \citet{green2011distances} uses the technique of {\rm H~}\kern 0.1em{\sc i} Self Absorption (HISA) assuming a flat rotation curve \citep{reid2009trigonometric,mcmillan2010uncertainty} to determine the kinematic distances towards 442 6.7GHz methanol masers. The distances reported in \citet{urquhart2013atlasgal} is a modified version of those presented in \citet{green2011distances} in that the Galactic rotation curve of \citet{brand1993velocity} is assumed to account for the significant variations between the model-derived tangent velocities and the empirically derived values from the {\rm H}\kern 0.1em{\sc i} termination velocities in the fourth quadrant. We have made use of the distances of \citet{urquhart2013atlasgal} instead of those in \citet{green2011distances} for our analysis. \citet{green20176} presents the distance towards an additional 202 methanol maser sources using the HISA method. 

Recently, the distances to a large number of high-mass star forming regions have been estimated using trigonometric parallax with Very Long Baseline Interferometry (e.g. \citealt{reid2014trigonometric} and references therein). Since high-mass star forming regions are expected to be good tracers of spiral arms in galaxies, \citet{reid2016parallax} suggested use of a Bayesian approach to assign sources to spiral arms based on their location and radial velocities and comparing with spiral arm signatures as seen in CO and {\rm H}\kern 0.1em{\sc i} surveys. \citet{reid2016parallax} claim that the use of this method should significantly improve the accuracy and reliability of distance estimates to sources that are good tracers of spiral structure. \citet{green20176} apply this technique to estimate distances to the entire 972 methanol maser sources that are catalogued to date with the consideration that 6.7~GHz methanol masers are mostly associated with high-mass star formation, which trace spiral structure.

We have used the new distances of \citet{green20176} to test whether any of the results derived in the previous sections are significantly altered. We first test whether the distances adopted from earlier references are significantly different from that of \citet{green20176} (methods 1 and 2 respectively hereafter). We compared the two sets of distances with a t-test, which yielded a t-value of 1.27 with a significance value of 0.20. Thus it can be inferred that the distance values calculated using these two methods are not significantly different. This is confirmed by examining the mean and median distances from the two methods -- while method 1 gives a mean and median distance of 8.03~kpc and 8.55~kpc, the values from method 2 are 8.07~kpc and 8.25~kpc respectively. A histogram showing the distribution of distance values are shown in  Fig.~\ref{fig:distcomp}. 

We further analysed the differences in physical parameters computed using distances from methods 1 and 2. The left panel of Fig.~\ref{fig:Mandmr} shows a comparison of the mass distribution from method 1 (green hatched histogram) and method 2 (solid red histogram). As expected from the differences in the distance distribution, the mass distribution from the two methods are slightly different. However, the overall statistics are similar -- while method 1 gives a mean and median mass of 1570~M$_\odot$ and 930~M$_\odot$, the respective values obtained by distances from method 2 are 1785~M$_\odot$ and 925~M$_\odot$. A t-test between the two mass distributions gives a t-value of 0.71 with a significance value of 0.48. Thus, there is no significant difference in the distribution of masses computed using distances from the two methods.

The right panel of Fig.~\ref{fig:Mandmr} shows the mass-radius relation computed using distances from method 2. With method 2, 287 out of 320 sources satisfy the Kauffmann criterion for potential to form massive stars. This is comparable to method 1 wherein 295 out of 320 sources satisfy the Kauffmann criterion. Thus, although the distances from method 1 are different from that of method 2, the overall statistical properties of sources associated with 6.7~GHz methanol masers are similar between the two methods. We are thus unable to make any distinction regarding accuracy of distances when considering the entire sample statistically.

\section{CONCLUSION}\label{sec5}
We have constructed SEDs from 870~$\mu$m to 70~$\mu$m for 320 6.7~GHz methanol masers using data from the ATLASGAL and Hi-GAL surveys. The SEDs from 870~$\mu$m to 160~$\mu$m were fit with single component grey body models. We observe a mean dust temperature of 22~K confirming the later evolutionary stage of the maser sources in comparison to infrared dark clouds, with some sources showing temperatures as high as 48~K. Almost 92\% of the methanol maser sources satisfy the Kauffmann criterion for potential to form massive stars. A comparison of the mass-luminosity diagram of the sample with simple evolutionary tracks from the turbulent core model suggest that most methanol masers are associated with massive young stellar objects with over 90\% in early evolutionary stages of accelerating accretion. However, there also appears to be a small population of sources that are likely to be associated with intermediate or low-mass stars suggesting that the association between high-mass star formation and methanol maser emission is not exclusive. We have also compared the physical parameters inferred from the use of the new Bayesian method of distance computation with that of the traditional kinematic distances and found no statistical differences in the same.

\section{ACKNOWLEDGMENTS}
We thank the referee whose comments helped in improving the paper. This research has  made  use of NASA's Astrophysics Data System and VIZIER service operated at CDS, Strasbourg, France. The \large matplotlib package \citep{Hunter2007} for \large python was used for making plots. We have used the online freemium academic writing environment Overleaf available at: https://www.overleaf.com/.

\bibliographystyle{mnras}
\bibliography{paper_google_scholr}


\bsp	
\label{lastpage}
\end{document}